\documentclass[12pt,a4paper]{article}
\usepackage{amsmath}
\usepackage{graphicx}
\usepackage{graphicx,amssymb,amsfonts,latexsym,amsmath,amsthm,times}
\usepackage{epsfig}
\usepackage{float}
\usepackage{fancyhdr}
\usepackage[english]{babel}
\usepackage{color}
\usepackage{subfigure}

\begin{document}

\vspace*{1mm} 

 \begin{center}
{\Large \bf Turing Instability in an Economic-Demographic\\ 
 Dynamical System Can Lead to Pattern Formation\\ \vspace*{2.5mm}
 on Geographical Scale
 }
 \end{center}

	\vspace*{12mm}

	\centerline{\large Anna Zincenko $^a$, Sergei Petrovskii $^a$\footnote{Corresponding
			author. E-mail: sp237@le.ac.uk}
		and Vitaly Volpert $^b$ }

	\vspace*{8mm}

	\centerline{$^a$ School of Mathematics \& Actuarial Science, University of Leicester}
        \centerline{Leicester, LE1 7RH, UK}
	
\vspace*{5mm}	
	
	\centerline{$^b$ Universit\'{e} de Lyon, Universit\'{e} Lyon1, CNRS UMR 5208 Institut
		Camille Jordan} \centerline{F-69200 Villeurbanne Cedex, France}

\vspace*{13mm}

 \begin{center}
{\bf Abstract.}
 \end{center}

Spatial distribution of the human population is distinctly heterogeneous, e.g.~showing significant difference in the population density between urban and rural areas. In the historical perspective, i.e.~on the timescale of centuries, the emergence of the densely populated areas at their present locations is widely believed to be linked to more favourable environmental and climatic conditions. In this paper, we challenge this point of view. We first identify a few areas at different parts of the world where the environmental conditions (quantified by the temperature, precipitation and elevation) are approximately uniform over thousands of miles. We then examine the population distribution across those areas to show that, in spite of the homogeneity of the environment, it exhibits a clear nearly-periodic spatial pattern. Based on this apparent disagreement, we hypothesize that there exists an inherent mechanism that can lead to pattern formation even in a uniform environment. We consider a mathematical model of the coupled demographic-economic dynamics and show that its spatially uniform, locally stable steady state can give rise to a periodic spatial pattern due to the Turing instability. Using computer simulations, we show that, interestingly, the emergence of the Turing patterns eventually leads to the system collapse.

\vspace*{13mm}


\newpage

\section{Introduction}

Fast growth of the global human population has long been regarded as a major challenge that faces the mankind \cite{Malthus,Ehrlich68,UN02,UN17}. Presently, this challenge is becoming even more serious than before, in particular because many natural resources are estimated to deplete before the end of this century. The increasing population pressure on the agriculture and on the ecosystems and the environment more generally is predicted to result in worldwide food and water shortages, pollution, lack of housing, poverty and social tension. The situation is exacerbated by the global climate change as considerable areas of lend are predicted to be flooded and hence taken out of human’s use. It is widely believed that, unless alternative scenarios of sustainable population growth and social development are identified and implemented, the mankind is likely to experience stagnation or even decline \cite{Meadows72}.

Population growth in time is complemented with the population dynamics in space. Population distribution over space is hugely heterogeneous for a variety of reasons, to mention the climate, the history, and the economy just as a few. The spatial heterogeneity may result in significant migration flows that in turn can have a significant feedback on the local demography and the population growth.
On a smaller scale of individual countries and states, understanding of the factors affecting the population distribution in space is needed to ensure adequate development of infrastructure, transport, and energy network. Poorly informed decisions are likely to result in overcrowding and social problems in urban areas and/or lower quality of life in rural neighbourhoods.

\begin{figure}[!t]
\includegraphics[scale=0.7]{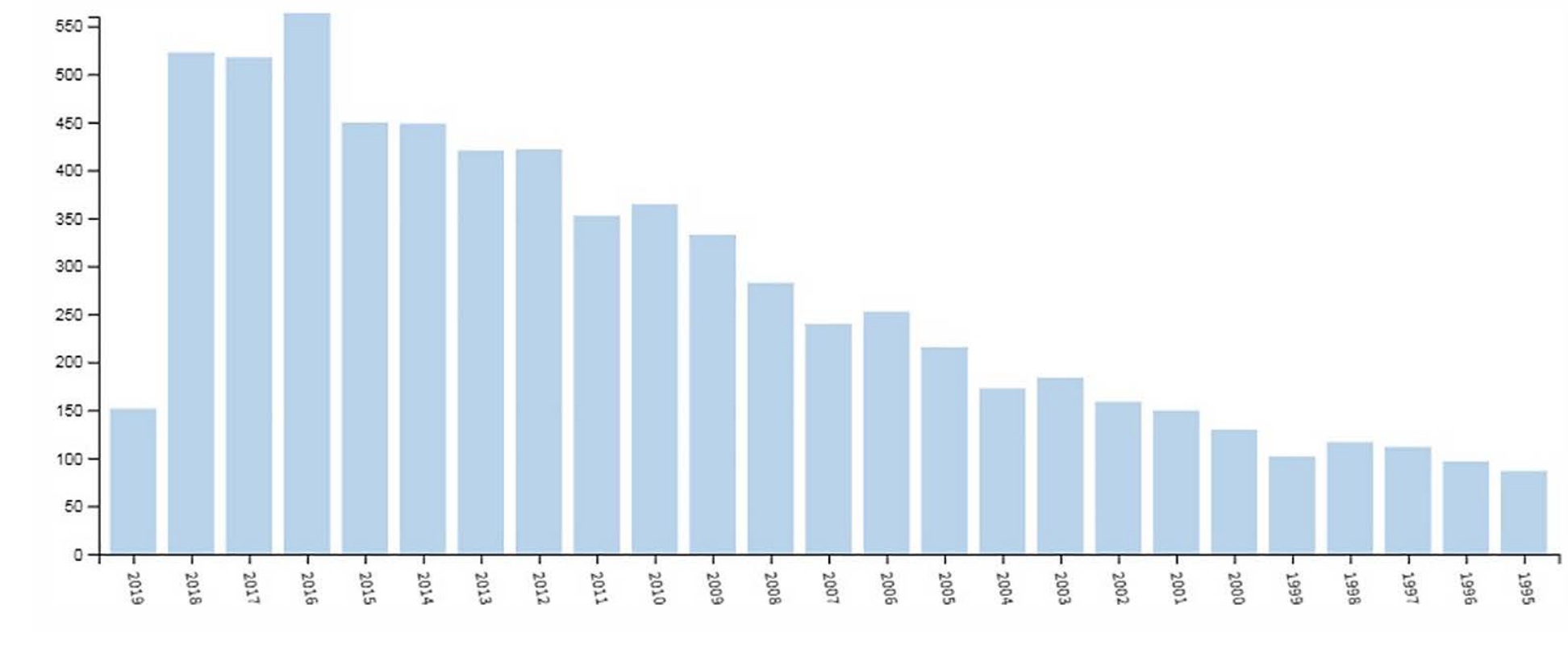}
\centering\caption{Annual number of publications over the last twenty five years dealing with mathematical modelling of demographic processes. Data are imported from the Web of Science. Note that time is increasing from right to left. The leftmost column shows the number of publications by mid-April 2019.}
\label{pubstats}
\end{figure}

Identification of scenarios of sustainable population growth and social development on various spatial and temporal scales requires good understanding of the relevant processes and mechanisms that affect both the population growth and the population distribution. Arguably, such understanding is unlikely to be achieved without a well-developed theory and the corresponding mathematical/modelling framework. Indeed, mathematical models of human population dynamics (e.g.~see \cite{Coale72,Impagliazzo85}) has a long history dating back to the 17th century \cite{Graunt}. Over the last few decades, the need for an adequate and efficient mathematical theory of the human population dynamics
has been reflected by a steady growth in the number of studies where problems of demography were considered using mathematical models, tools and techniques; see Fig.~\ref{pubstats}.

 In this paper, we use mathematical modelling to address the phenomenon of heterogeneous spatial population distribution.
Heterogeneity of geographical features (mountains, forests, rivers, etc.) and natural resources (e.g.~coal, iron and copper ore) are commonly accepted as factors leading to the demographic and economic heterogeneity. However, the question  that we ask here is -- is this natural heterogeneity the only underlying cause, or can there be another and perhaps more general principle responsible for emergence of heterogeneous population distribution?

In order to answer this question, we first revisit available data on the population density over a few areas in different parts of the world to show that, in all cases, the population distribution exhibits a clear nearly-periodic spatial pattern in spite of the fact that the environmental conditions are relatively uniform.
 We then consider a novel model of coupled economic-demographic dynamics in space and time and endeavour to use it to simulate the spatial population distribution. The model consists of two coupled partial-differential equations of reaction-diffusion type.
   We show that the emergence of spatial patterns appears to be possible as a result of Turing instability.
By relating the model predictions to the data on the human population density, we argue that the heterogeneous population distribution observed across different countries in different continents may have been caused by endogenous rather than exogenous factors, i.e.~may have appeared due to intrinsic Turing instability of the corresponding economic-demographic dynamical system.

\section{Real-world examples}\label{sec:realdata}

In many countries, the population distribution over space is distinctly heterogeneous, e.g.~urbanized areas with a high population density alternate with rural areas with a low population density. Apparently, spatial variation in geographical and climatic factors can play a significant role in shaping the population distribution. Since our main hypothesis in this paper is the existence of a dynamical mechanism that may lead to the formation of heterogeneous population distribution regardless of the geographical heterogeneity, in our search for the real-world examples we focus on the cases where the environment may be regarded as relatively uniform. The environmental properties that we consider here as proxies for the environmental heterogeneity are the elevation, the annual mean temperature, and the annual mean precipitation. A brief overview of the several relevant cases is given below.

\subsection{Canadian Southern Region}

\begin{figure}[!t]
\centering
 \subfigure[]{\includegraphics[scale=0.54]{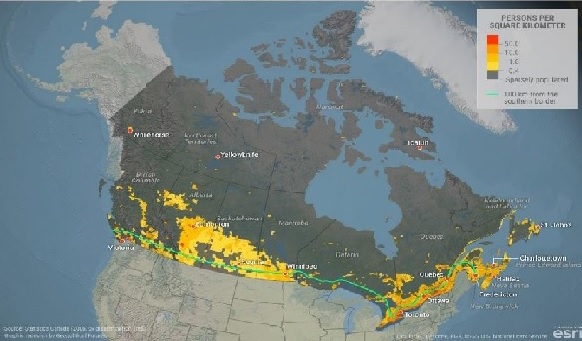}}\hspace{5mm}
 \subfigure[]{\includegraphics[scale=0.55]{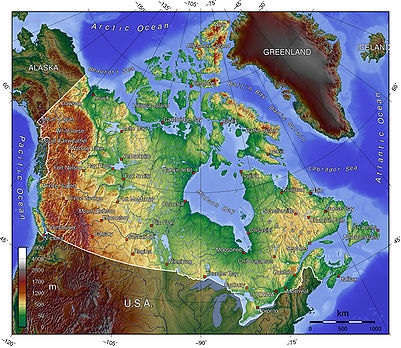}}\vspace*{4mm}
 \subfigure[]{\includegraphics[scale=0.6]{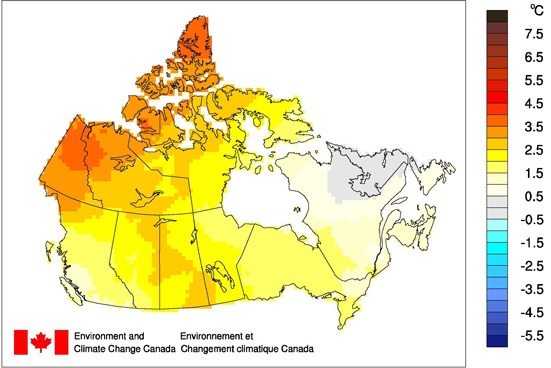}}\hspace{5mm}
 \subfigure[]{\includegraphics[scale=0.27]{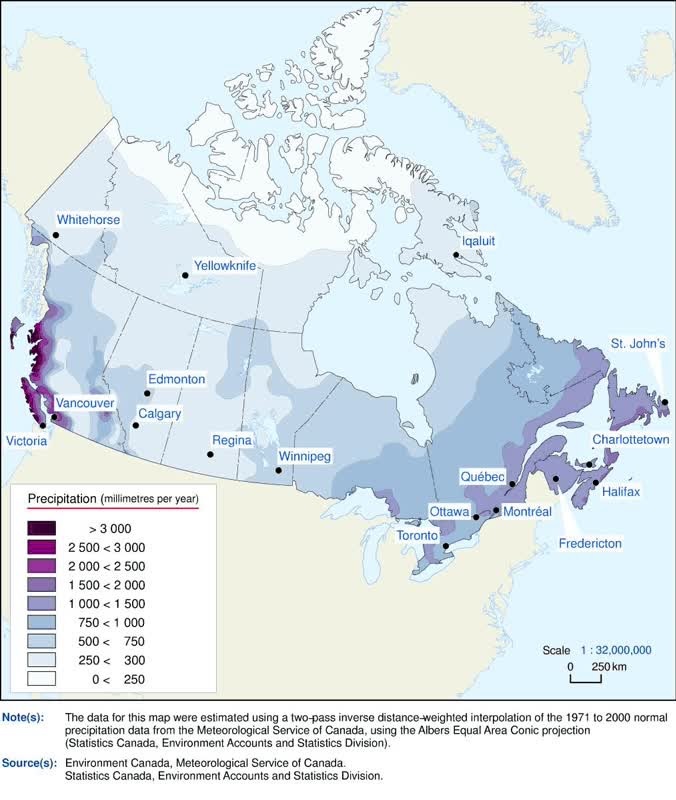}}
\caption{\small (a) Population distribution in Canada. Adapted from \cite{CanadaYearBook}. A strongly heterogeneous, `patchy' structure is readily seen. The green line indicates the border of the 160km wide strip where the environmental conditions are approximately homogeneous.
(b) Geographic map of Canada showing the elevation \cite{TopoCanada}. One can see that, to the east of the Rocky Mountains, the elevation along the Canada-USA border is approximately uniform. (c) Annual mean temperature map of Canada in 2016 \cite{ClimateCanada}. It is readily observed that the temperature does not vary much along the southern border. (d) Precipitation map of Canada \cite{EnvironmCanada}. The amount of precipitation does not vary  much along the border, except for the extreme West.}
 \label{GCanada}
\end{figure}

Canada is a scarcely populated country and the majority of Canadian population live in the narrow band (approx.~160 km) along the USA border, see Fig.~\ref{GCanada}a. The distribution of the environmental properties across the country is highly heterogeneous, in particular in the South-North direction, ranging from temperate climate in the South to the rather extreme polar climate in the North. However, the magnitude of climatic variation in the East-West direction is much smaller (see Fig.~\ref{GCanada}c), at least over the span between the Atlantic coast and the Rocky Mountains where the annual mean temperature varies just within 2-3$^{\circ}$C (contrary to about 20$^{\circ}$C in the South-North direction). A similar observation applies to the elevation and the annual mean precipitation; see Figs.~\ref{GCanada}b and ~\ref{GCanada}d, respectively.

\begin{figure}[!t]
\hspace{-3mm}\subfigure[]{\includegraphics[scale=0.54]{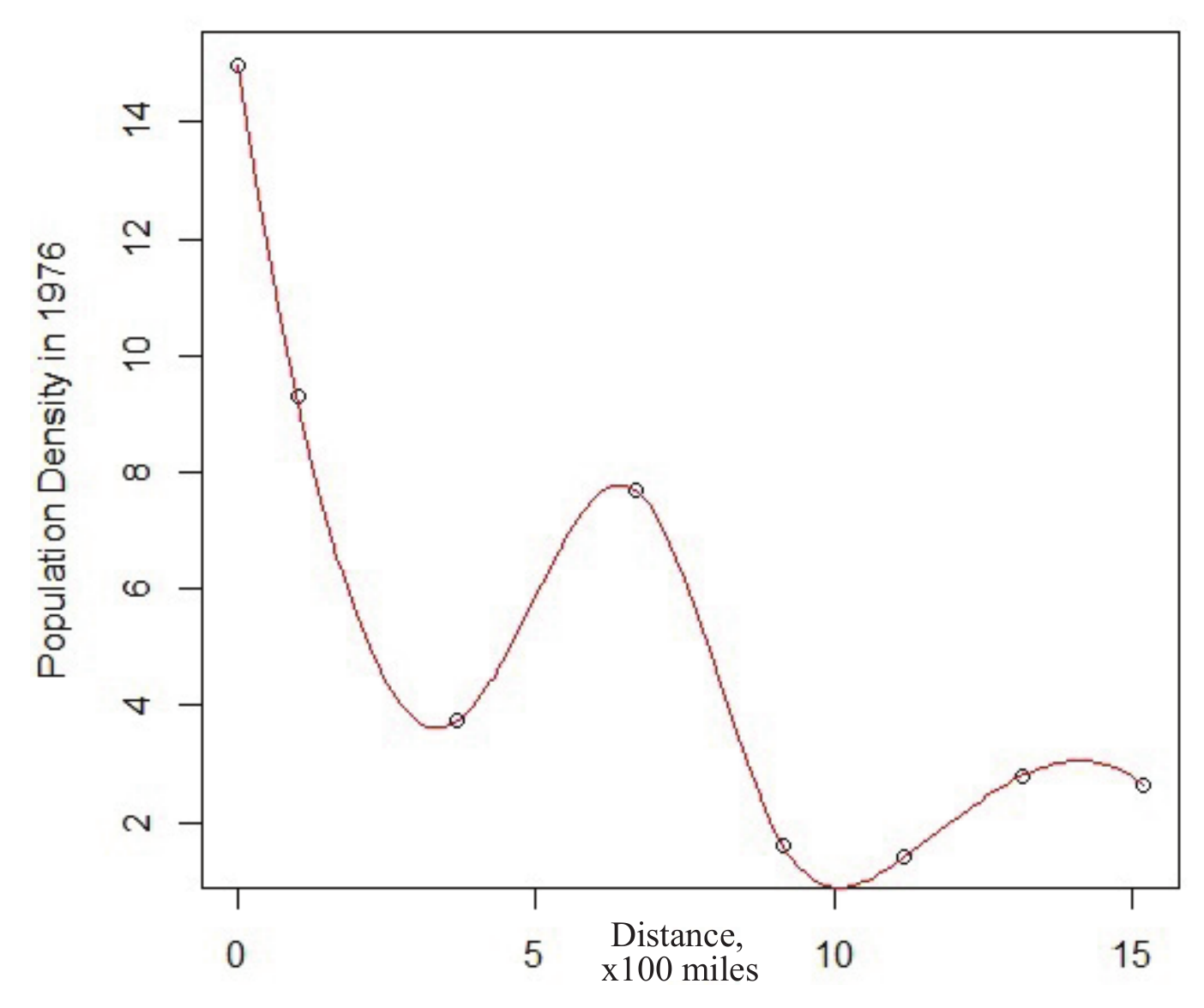}}\hspace{7mm}
             \subfigure[]{\includegraphics[scale=0.48]{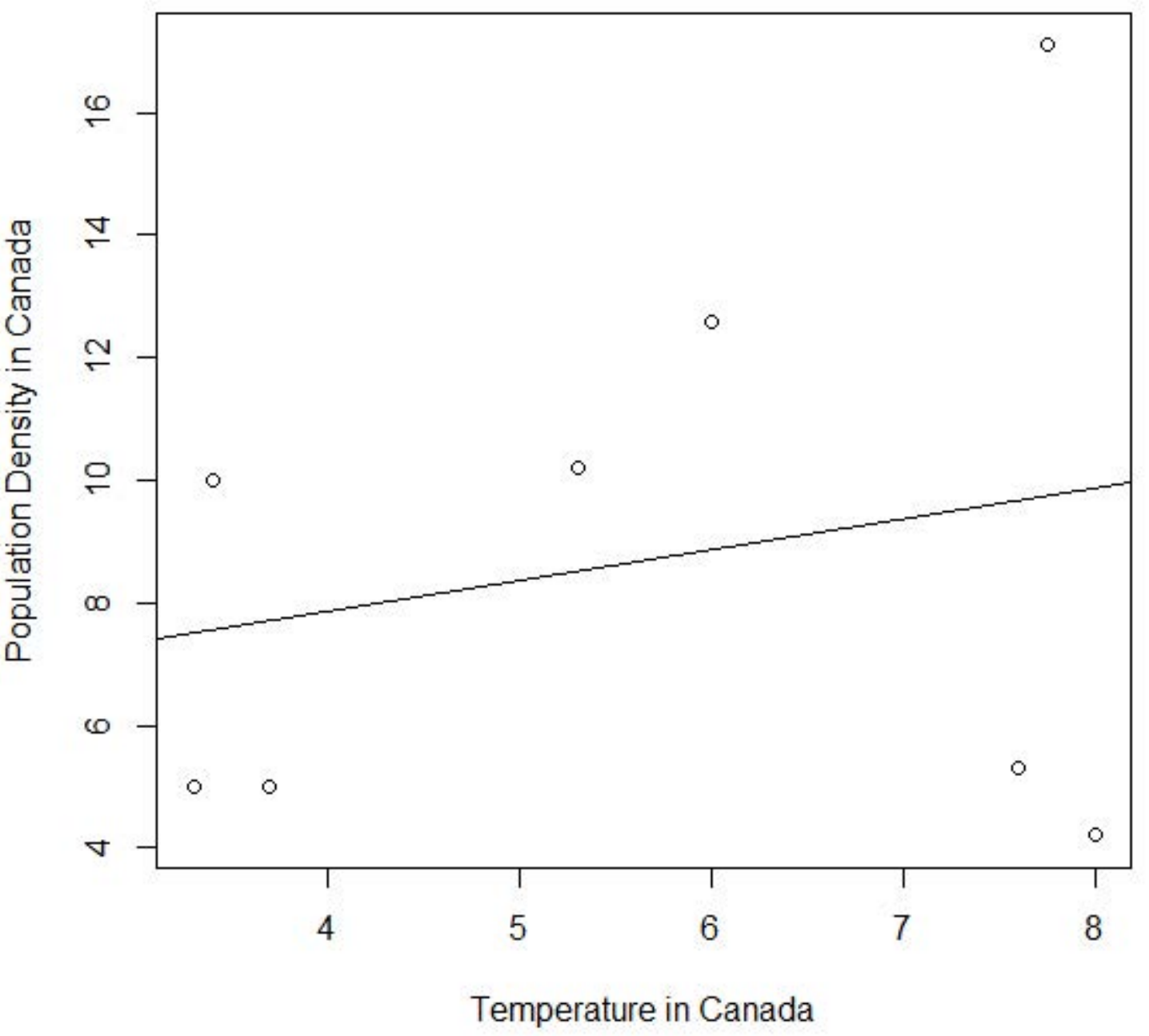}}\\
\hspace{3mm}\subfigure[]{\includegraphics[scale=0.5]{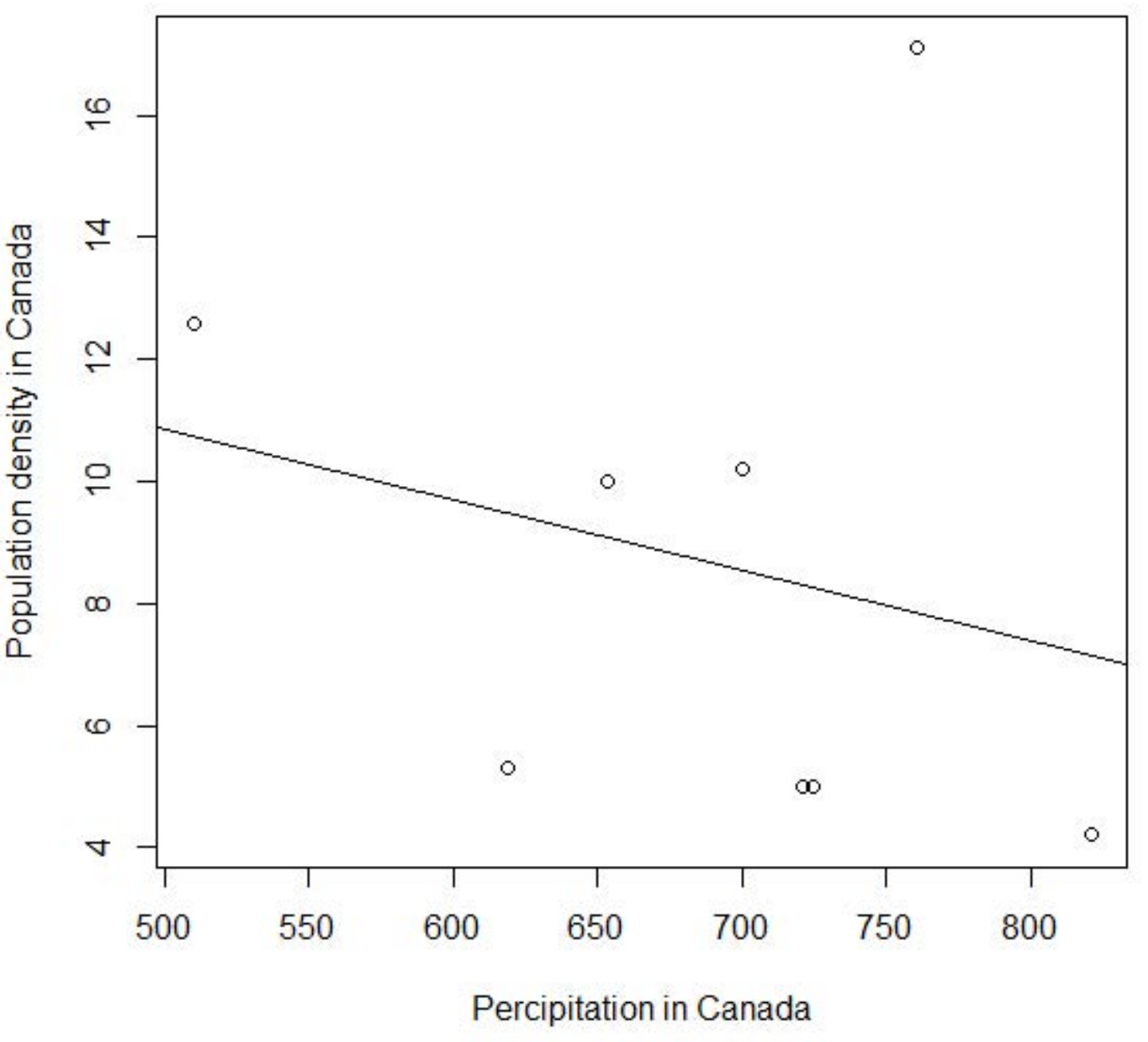}}\hspace{7mm}
            \subfigure[]{\includegraphics[scale=0.5]{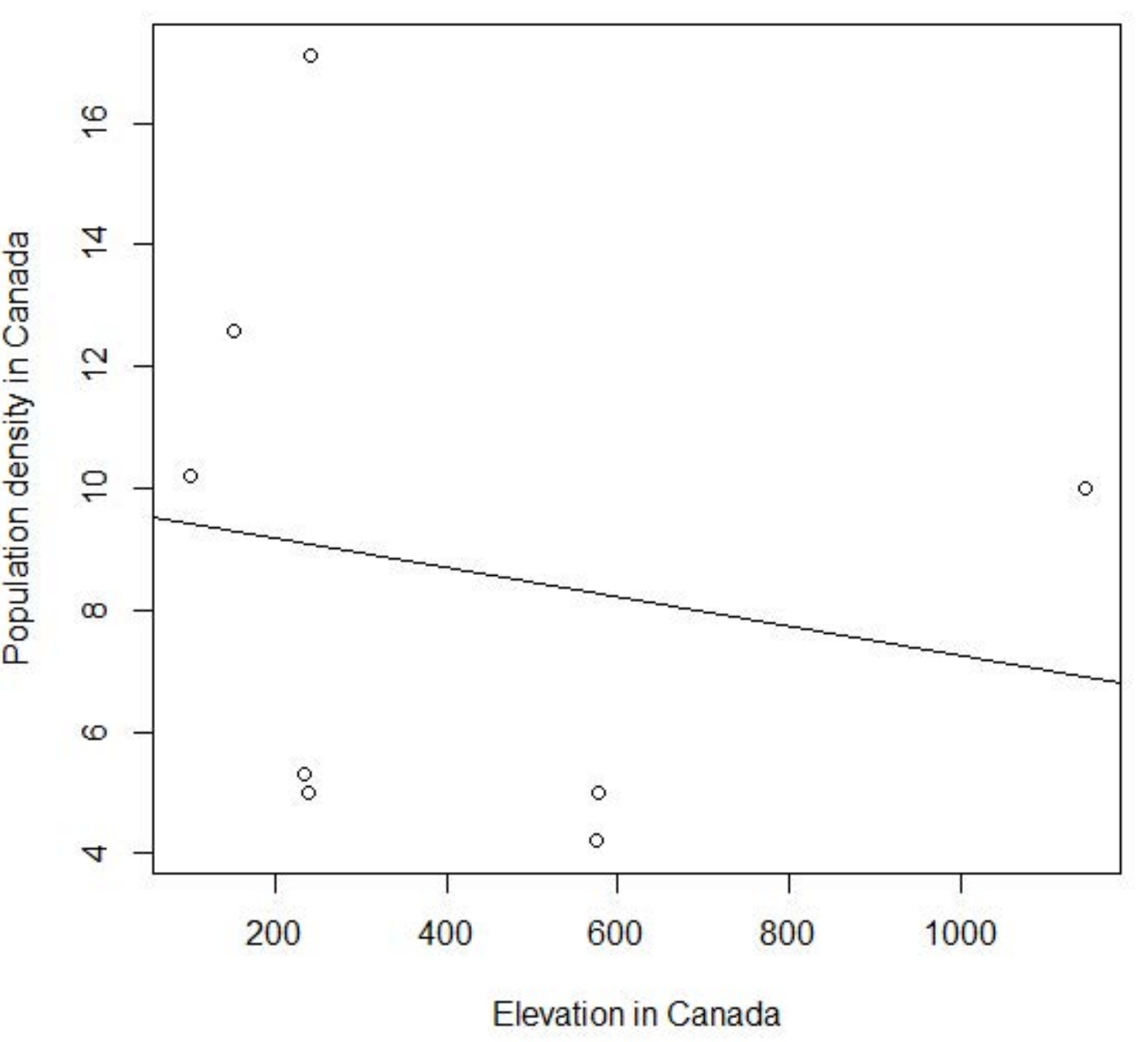}}
\caption{{\small (a) Population density vs space (shown in hundreds miles) in Canada in 1976 in the East-to-West direction in the 160 miles wide corridor along the southern border (see the green line in Fig.~\ref{GCanada}a). (b) Population density (axis $Y$) vs the annual mean temperature (axis $X$) in Canada. (c) Population density (axis $Y$) vs the annual mean precipitation (axis $X$) in Canada. (d) Population density (axis $Y$) vs the elevation (axis $X$). In all cases shown in panels (b-d), the best-fitting straight line is drawn by maximizing $R^2$; for details, see Table \ref{tabR2ind}.}}
 \label{Pictureddc}
\end{figure}

We now focus our analysis on the curved narrow corridor along the border (see the green line in Fig.~\ref{GCanada}a) where the environmental conditions are relatively uniform but the population distribution is not.
Figure \ref{Pictureddc}a demonstrates how the population density inside the band varies in space along the East-West direction. Interestingly, the distribution exhibits three maxima with approximately equal spacing of 700 miles. We therefore regard it as a periodic spatial distribution. Note that this pattern is persistent over time: a similar periodic-like structure is observed for different years (not shown here for the sake of brevity) starting from at least late 19th century.

In order to reveal how strong is the effect of environmental properties on the population distribution, we now perform the pairwise correlation analysis between the population density and each of the three environmental factors that we consider here. The results are shown in Figs.~\ref{Pictureddc}b-d and the corresponding values of the coefficient of determination $R^2$ \cite{Grantz90} are given in Table \ref{tabR2ind}. We readily observe that the obtained values of $R^2$ are quite small, hence only a small proportion of the variance in the population density can be explained by 
the environmental factors \cite{Grantz90}.

\begin{table}[!t]
	\caption{Values of the coefficient of determination $R^2$ \cite{Grantz90} for the best-fitting linear function between the population density in the three considered countries and the three environmental properties.}
	\label{tabR2ind}
	\begin{center}
		\begin{tabular}{|c|c|c|c|}
\hline
			       & Elevation,  & Annual mean           & Annual mean  \\
                   &  meters     & temperature, $^\circ C$ & precipitation, mm \\
\hline
			Canada & $-0.14$ & $-0.15$  & $-0.09$  \\
\hline
		 Australia & $-0.14$ & $-0.04$  & $-0.06$  \\
\hline
		  Mongolia & $-0.08$ & $-0.08$  & $-0.04$ \\
\hline
		\end{tabular}
	\end{center}
\end{table}

\begin{figure}[!b]
\centering
 \subfigure[]{\includegraphics[scale=0.51]{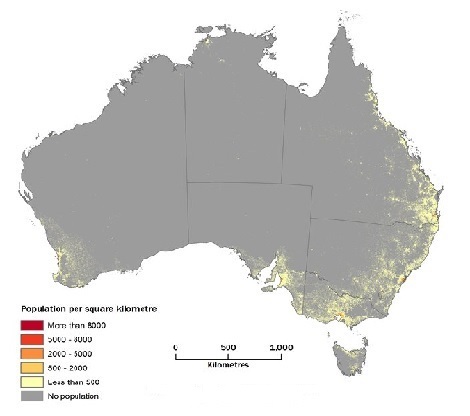}}\hspace{7mm}
 \subfigure[]{\includegraphics[scale=0.51]{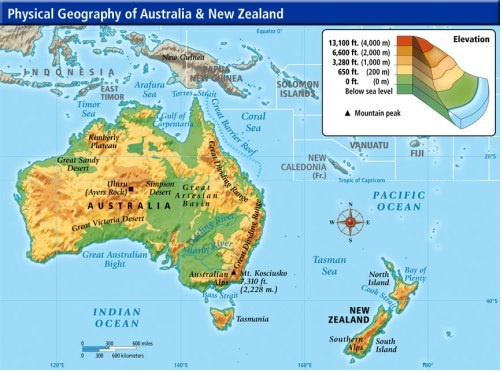}}\vspace*{3mm}
 \subfigure[]{\includegraphics[scale=0.32]{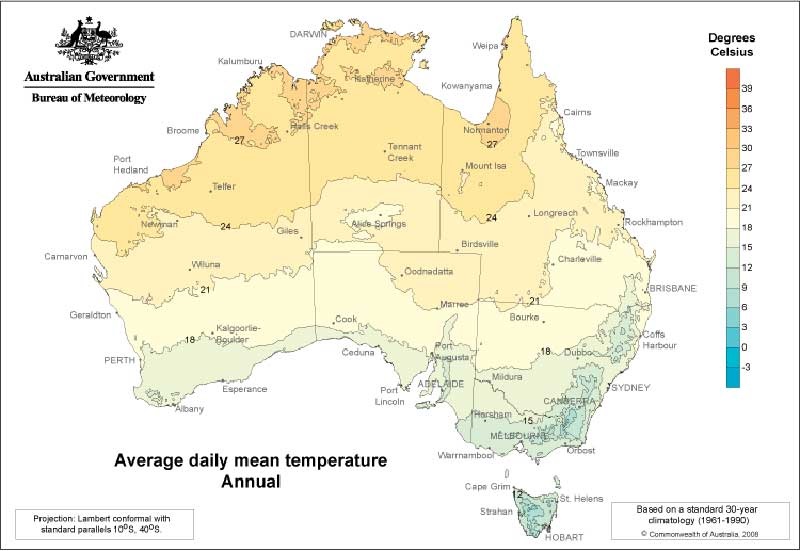}}\hspace{5mm}
 \subfigure[]{\includegraphics[scale=0.34]{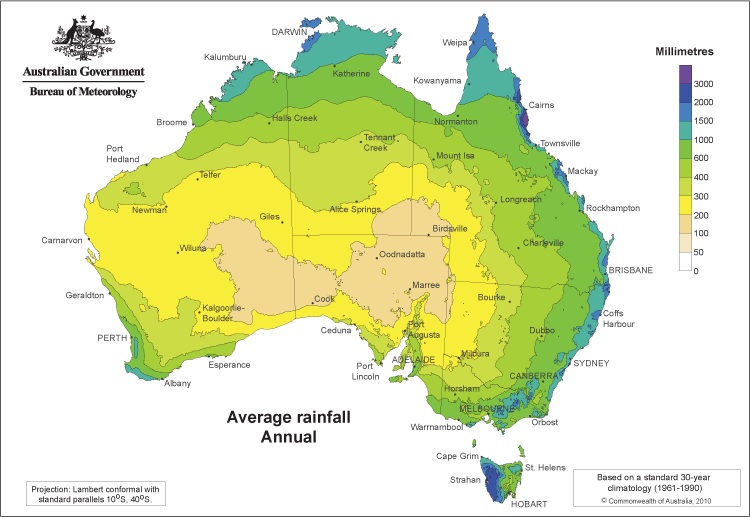}}
\caption{{\small (a) Population distribution in Australia in 2011. It can be seen that the population is concentrated in the Southeast. From \cite{AustralPopDens}. (b) Geographic map of Australia showing the elevation. From \cite{AustralElevat}. (c) Annual mean temperature map based on 30 years observations, 1961-1990. From \cite{AustralTemper}. (d) Annual precipitation map based on 30 years observations, 1961-1990. From \cite{AustralPrecipit}.}}
 \label{Australia}
\end{figure}

\subsection{South-Eastern Australia}

\begin{figure}[!b]
 \subfigure[]{\includegraphics[scale=0.54]{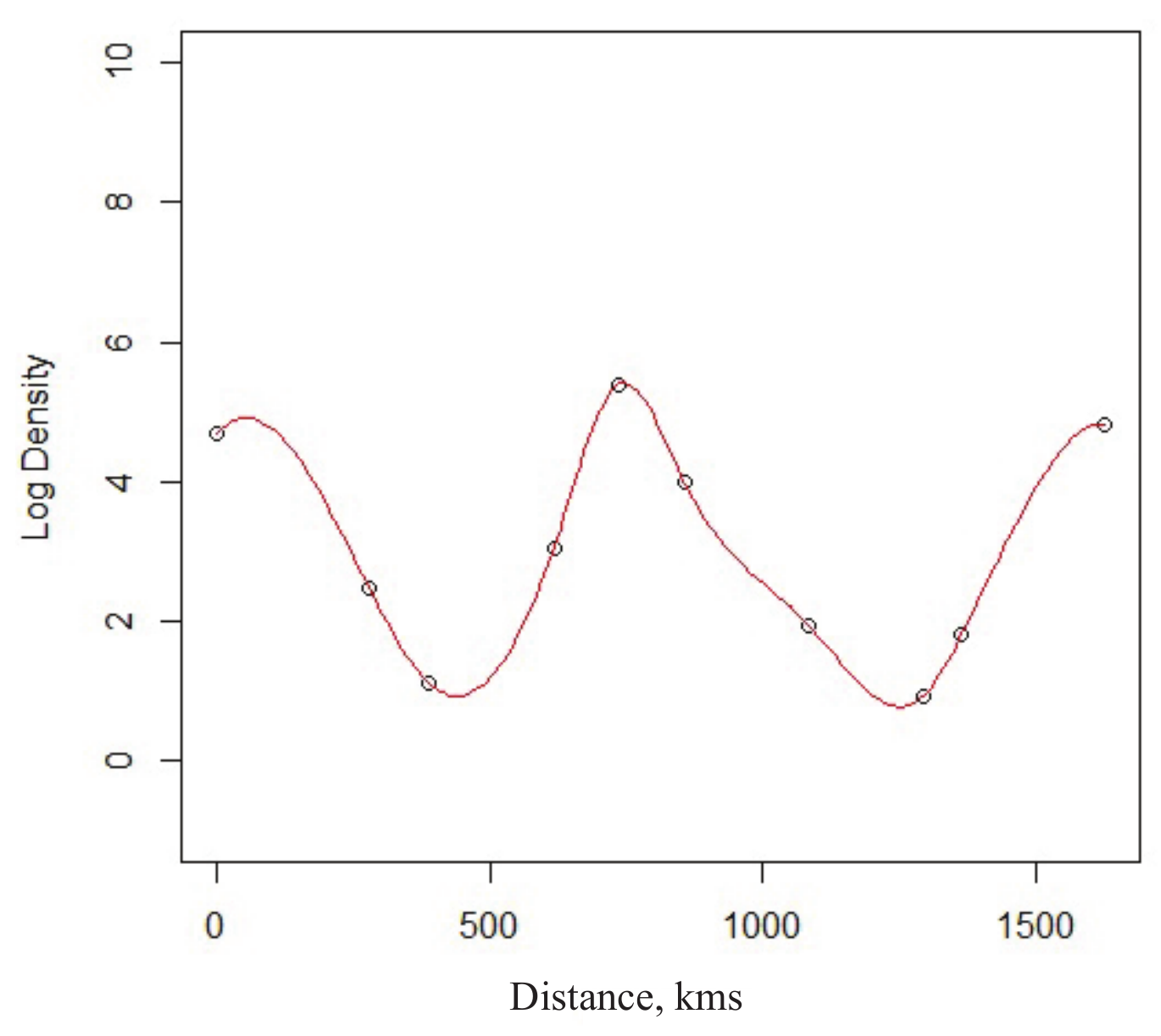}}\hspace{6mm}
 \subfigure[]{\includegraphics[scale=0.49]{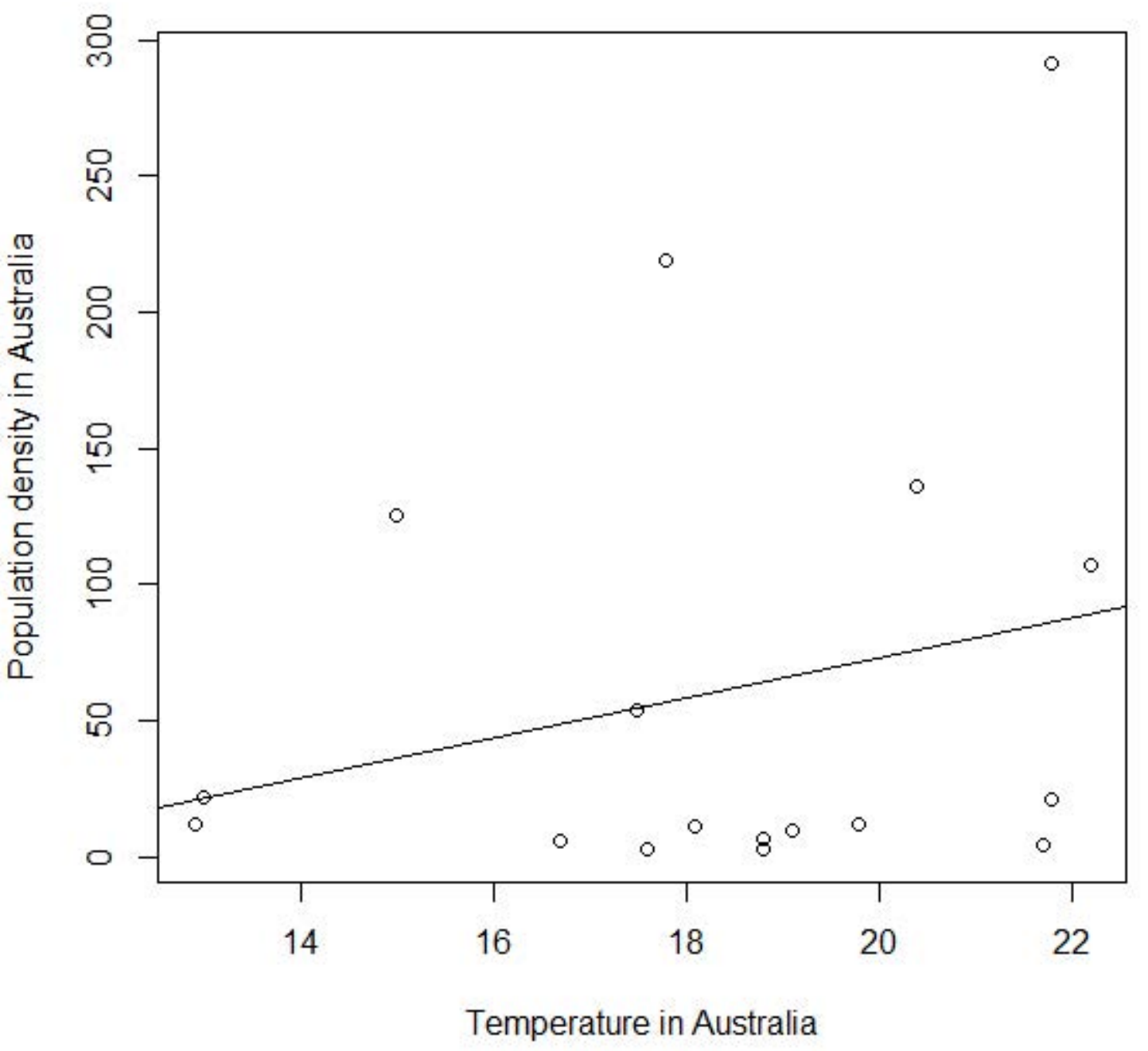}}\\
 \subfigure[]{\includegraphics[scale=0.5]{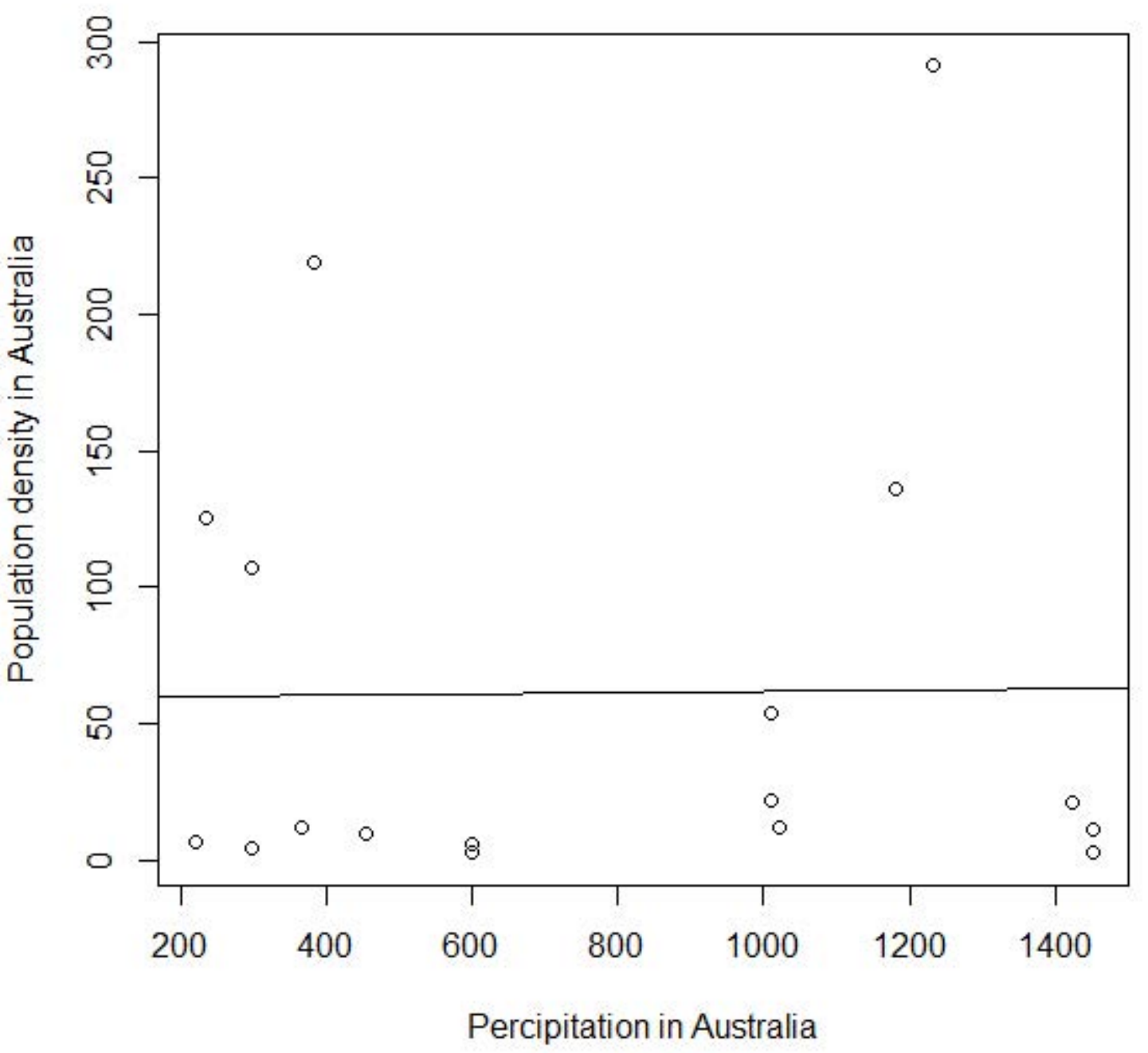}}\hspace{6mm}
 \subfigure[]{\includegraphics[scale=0.5]{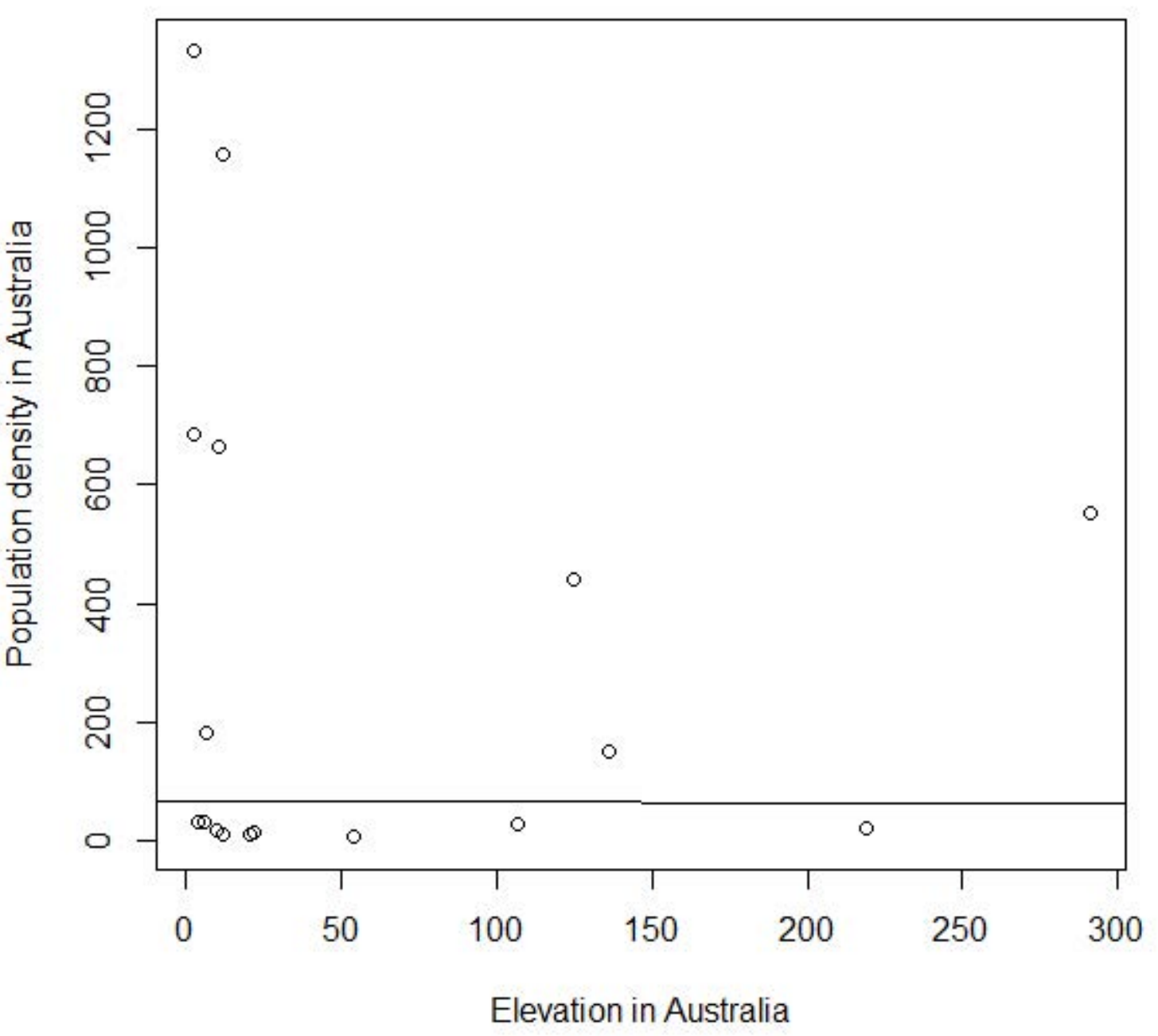}}
\caption{{\small (a) Population density vs space inside the 400 km wide stripe along the coast of Tasmanian Sea and Bass strait in the southeast Australia. Axis $X$ shows the distance in km from Brisbane along the coastline. (b) Population density (axis $Y$) vs the annual mean temperature (axis $X$) in Australia. (c) Population density (axis $Y$) vs the annual mean precipitation (axis $X$). (d) Population density (axis $Y$) vs the elevation (axis $X$). In all cases shown in panels (b-d), the best-fitting straight line is drawn by maximizing $R^2$; for details, see Table \ref{tabR2ind}.}}
 \label{AAustralia}
\end{figure}

As another relevant example of the heterogeneous population distribution in an approximately uniform environment, we consider South-Eastern Australia. As well as Canada, Australia is a scarcely populated country, with most of the Australian population concentrated in three regions, i.e.~South-East, East, and South-West; see Fig.~\ref{Australia}a. The most densely populated area is the South-East. This area has the shape of a narrow strip (approx.~250 miles wide) along the coast of the Tasmanian sea and Bass strait. It appears that the climatic properties along this narrow strip such as precipitation and temperature are approximately uniform (see Figs.~\ref{Australia}b-d), e.g.~the variation in the annual mean temperature is just a few degrees (compared to more than 30$^{\circ}$C over the continent as a whole). The stripe includes the Great Dividing Range and the Australian Alps, which therefore accounts for a significant variation in the elevation.

In spite of the relatively uniform environment (apart from the elevation, its effect is discussed below), the population distribution along the strip is clearly heterogeneous with the population density varying several times between the more dense areas and the less dense ones; see Fig.~\ref{AAustralia}a.
Interestingly, it exhibits a nearly-periodic pattern where the three maxima are approximately equally spaced by about 700-850 kms.

An immediate intuitive explanation of the heterogeneous population distribution can be sought in the heterogeneity of the environmental properties. Correspondingly,
we look into the effect of the environmental factors more carefully by considering the correlation between each of the three factors chosen above and the population density. Figures \ref{AAustralia}b and \ref{AAustralia}c show the scatterplots of the population density in Australia vs the mean annual temperature and the mean annual precipitation, respectively. In both cases, the straight line shows the best-fitting of the data to maximize $R^2$; the corresponding values of $R^2$ are shown in Table \ref{tabR2ind}. It is readily seen that in both cases $R^2$ is quite small. We therefore conclude that the climatic variation is unlikely to be the factor that defines the spatial distribution of the population.

Now we recall that the study area includes the mountain ranges and exhibits considerable variation in the elevation. The question hence arises as to whether that can be a relevant factor. However, we first notice that the vast majority of the Australian population leaves at the elevation below 250 meters; see Fig.~\ref{AAustralia}d. We then perform the correlation analysis by looking for the best-fitting straight line in the scatterplot of the population density vs the elevation. The corresponding value of $R^2$ (see Table \ref{tabR2ind}) appears to be very small. We therefore rule out the elevation as
a factor affecting the heterogeneous spatial population distribution along the South-East coast of Australia.

\subsection{Mongolian Grassland}

\begin{figure}[!t]
 \subfigure[]{\includegraphics[scale=0.55]{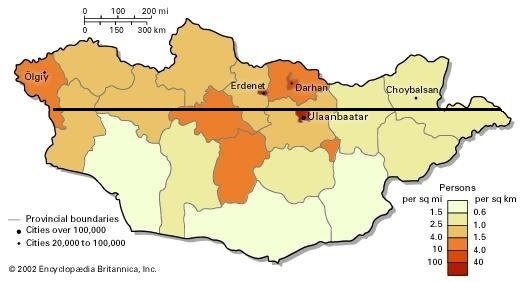}}
 \subfigure[]{\includegraphics[scale=0.65]{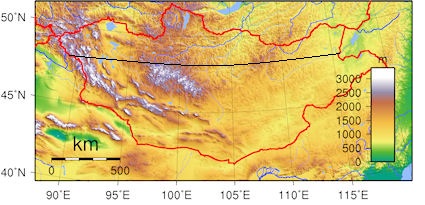}}\vspace*{3mm}
 \subfigure[]{\includegraphics[scale=0.2]{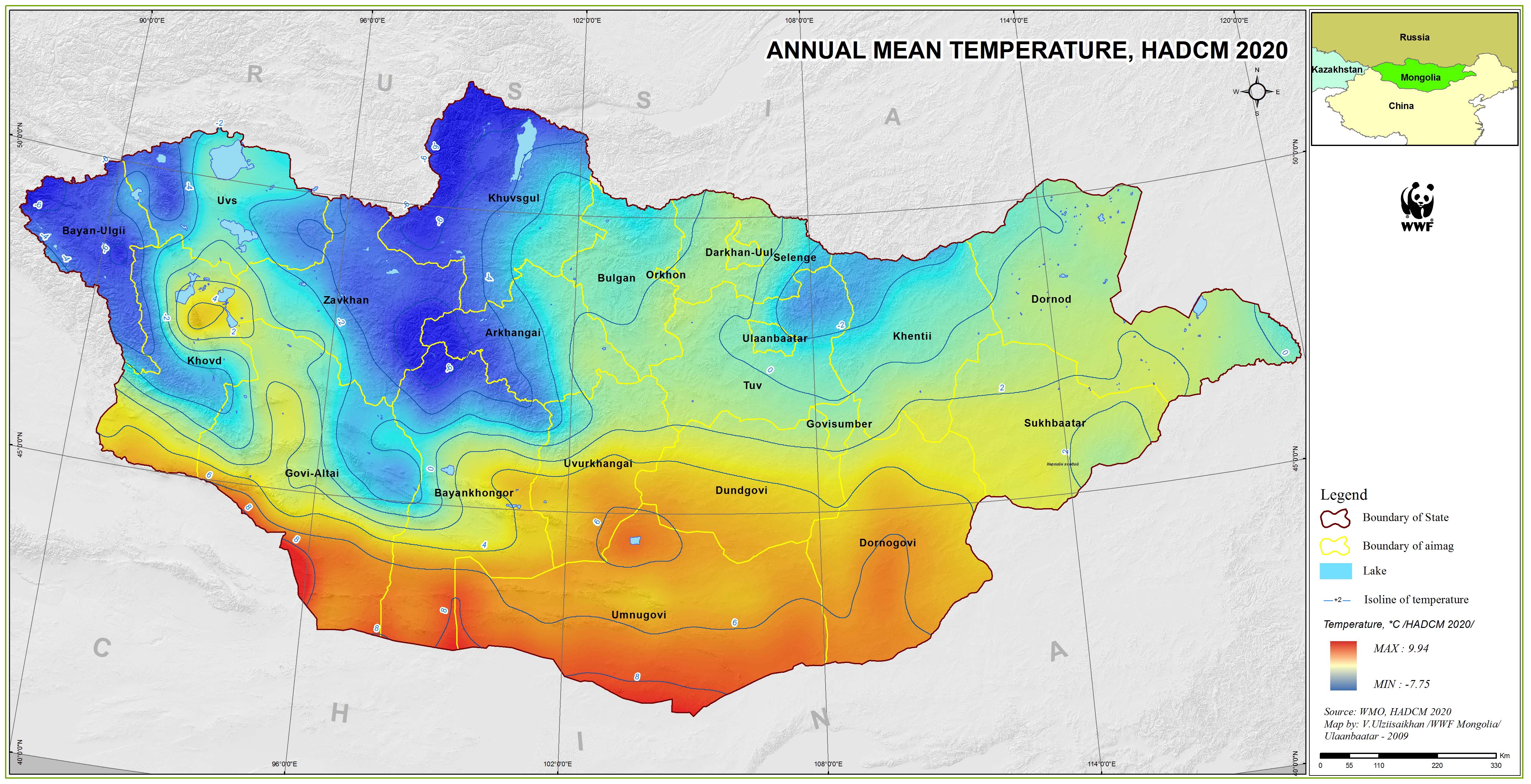}}\hspace{5mm}
 \subfigure[]{\includegraphics[scale=0.29]{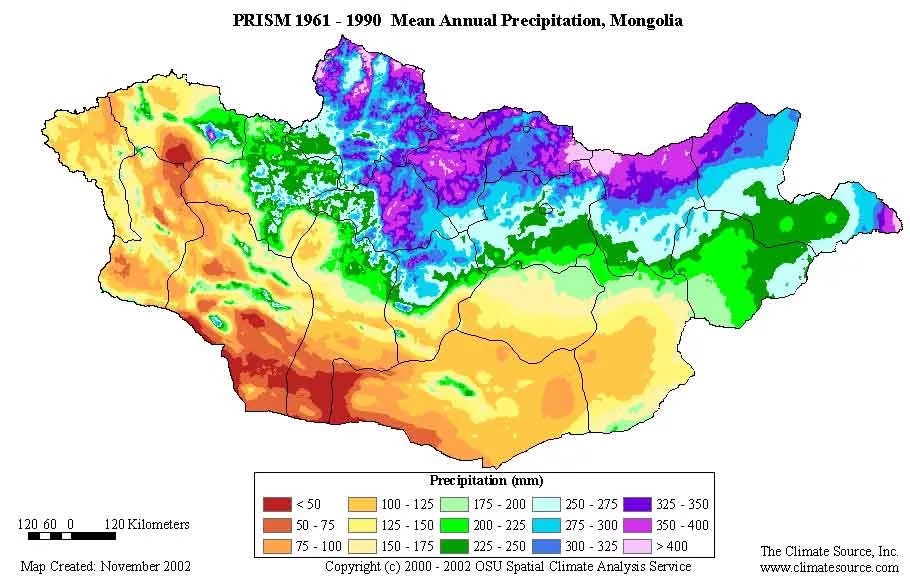}}
\caption{{\small (a) Spatial population distribution in Mongolia. The horizontal line shows the transect (along the 37.7$^{\circ}$ latitude) across the most densely populated areas. From \cite{MongoliaPopDens}. (b) Geographic map of Mongolia showing the elevation. From \cite{MongolElevat}. (c) Mean annual temperature. From \cite{MongolPrecip}. (d) Mean annual precipitation in Mongolia for the period 1961-1990. From \cite{MongolPrecip}.}}
 \label{Mongolia}
\end{figure}

\begin{figure}[!t]
 \subfigure[]{\includegraphics[scale=0.52]{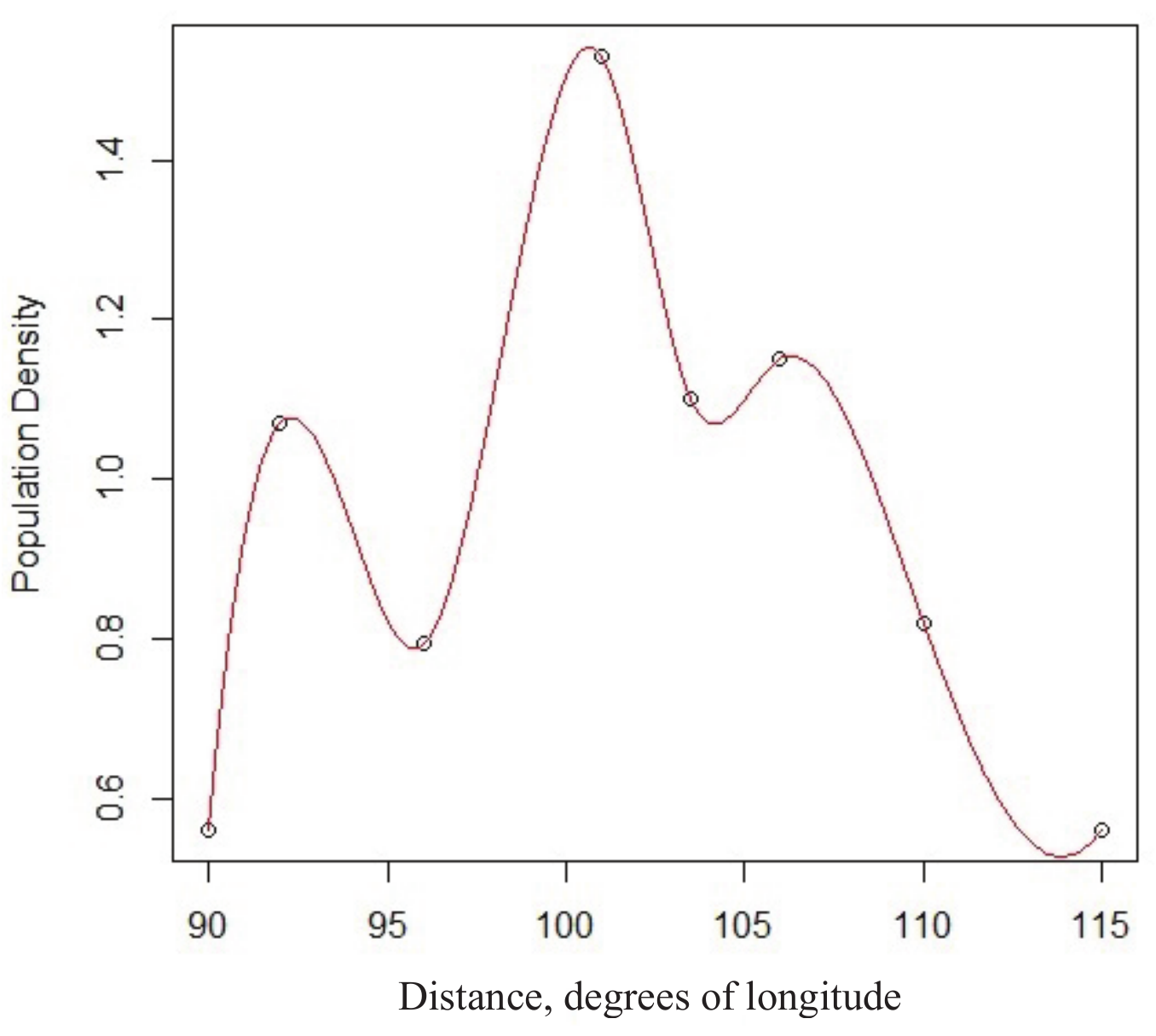}}\hspace{8mm}\subfigure[]{\includegraphics[scale=0.49]{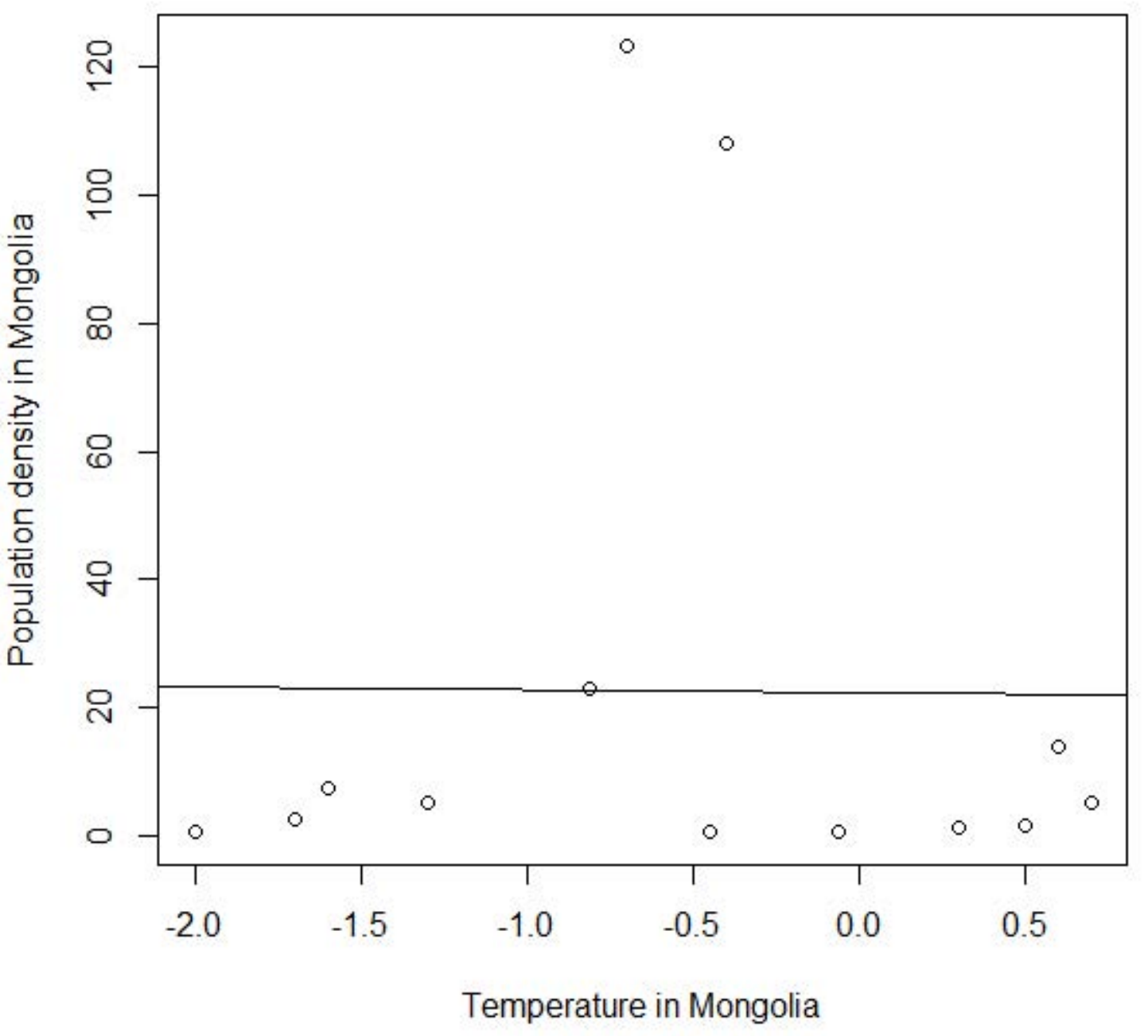}}\\
 \subfigure[]{\includegraphics[scale=0.5]{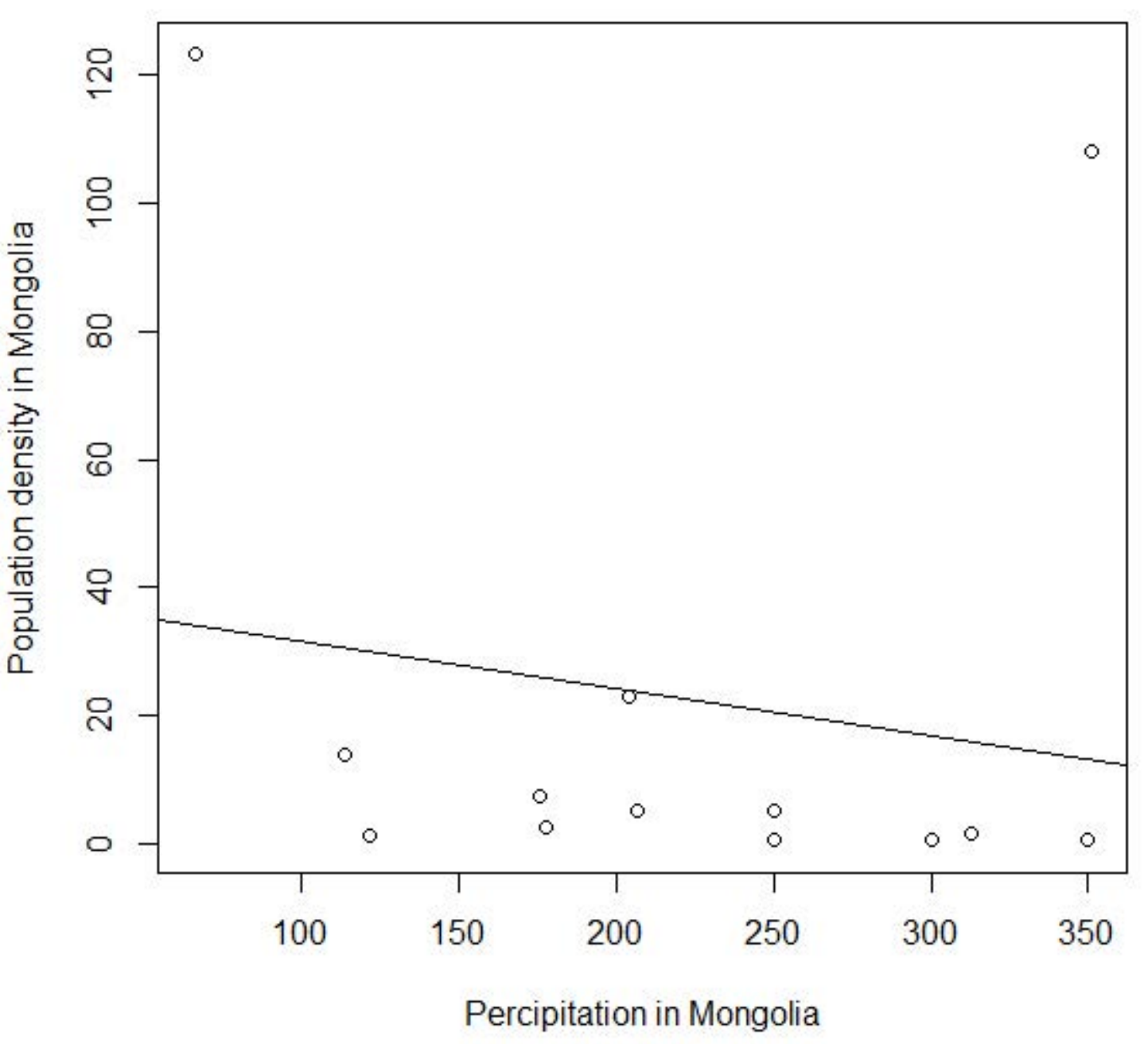}}\hspace{7mm}
 \subfigure[]{\includegraphics[scale=0.5]{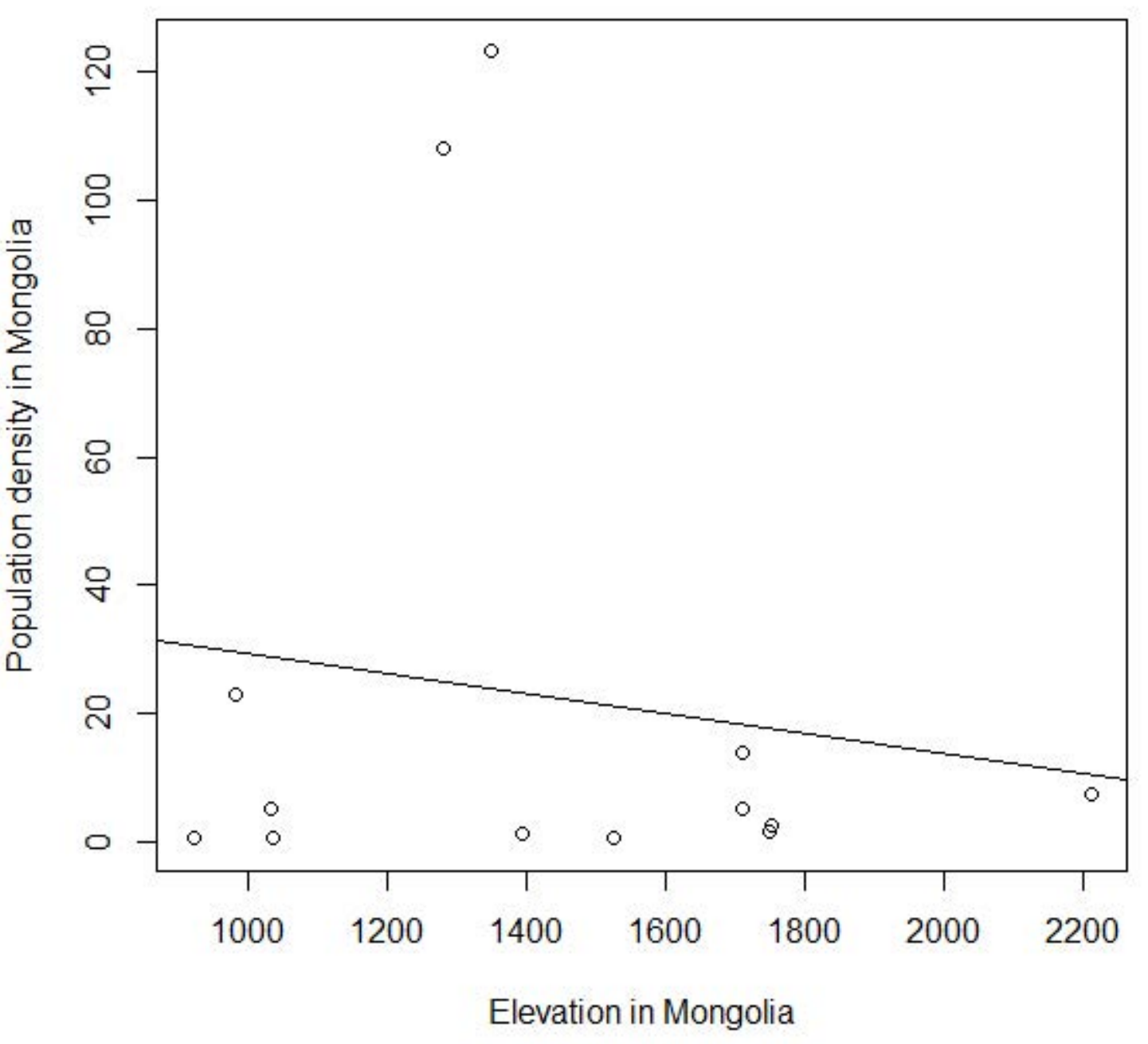}}
\caption{{\small (a) Population distribution across the central areas of the country in Mongolia. (b) Population density (axis $Y$) vs the mean annual temperature (axis $X$) in Mongolia. (c) Population density (axis $Y$) vs the mean annual precipitation (axis $X$) in Mongolia. (d) Population density (axis $Y$) vs the elevation (axis $X$). In all cases shown in panels (b-d), the best-fitting straight line is drawn by maximizing $R^2$; for details, see Table \ref{tabR2ind}.}}
 \label{MongoliaM}
\end{figure}

Mongolia, a Central-Asian country situated between China in the South and Russia in the North, has an elongated territory that extends from East to West for about 2400 km. It is the most sparsely populated country in the world. South of Mongolia is occupied by the Gobi Desert, which is barely populated at all due to the harsh climate and lack of resources.
 The majority of Mongolian three million population live in grasslands, which is located in the North of the country. In order to reveal the features of the spatial population distribution as is needed in the context of this study, we focus on the densely populated narrow corridor located along the latitude at 47.7 degrees North; see the black line in Fig.~\ref{Mongolia}a.
Interestingly, we readily observe that, as well as in the two previous cases, the population distribution in the East-West direction exhibits a periodic-like pattern (Fig.~\ref{MongoliaM}a). The three distinct peaks are separated by 700 and 900 kms intervals.

Variation of the environmental properties (cf.~Figs.~\ref{Mongolia}b-d) along the latitude is considerably less than in the North-South direction. However, it appears to be larger than it is in the cases of Canada and Australia, e.g.~the annual mean temperature varies over about 10$^{\circ}$C and the annual mean precipitation from 50 to 350 mm/(m$^2\cdot$year). Also the elevation varies over about 1500 meters, which is somewhat less than in Australia but larger than in Canada (where our analysis did not include the Rocky Mountains).

In order to reveal whether the variation of the environmental properties has any significant effect on the distribution of the population, we now perform the pairwise correlation analysis. The scatterplots of the population density vs the mean annual temperature, mean precipitation and the elevation are shown in Fig.~\ref{MongoliaM}b, c and d, respectively. The straight line is the best-fitting linear function; the corresponding values of $R^2$ are given in Table \ref{tabR2ind}. Apparently, the correlation between the population distribution and the environmental factors is very weak. We therefore conclude that the nearly-periodic pattern clearly seen in the population distribution is unlikely to be caused by the environmental conditions.

\subsection{Multiple linear regression}

In the above, we have shown that the spatial distribution of the population is unlikely to be affected, not to any considerable extent, by any single environmental property such as the mean annual temperature, mean precipitation or the elevation. However, generally speaking this does not rule out a possibility that a certain combination of those three factors may have a much stronger effect.  In order to check this possibility, we applied the multiple regression:
\begin{equation}\label{multivar}
y = a_{0}+\sum_{i=1}^3 a_{i} x_{i},
\end{equation}
where $y$ is the population density, $x_1$, $x_2$ and $x_3$ are, respectively, the average annual temperature, the annual precipitation and the elevation. Model (\ref{multivar}) was applied separately to the data for each of the three countries. The results are shown in Table \ref{tab}. We readily observe that the joint consideration of the three environmental factors does not real to a stronger correlation. Small values of $R^2$ indicate that variation of population density is only to a small amount explained by the geographical and climatic properties.

\begin{table}[!h]
\vspace*{5mm}
	\caption{Parameters of the linear model (\ref{multivar}) corresponding to the best fitting of the data obtained by maximizing $R^2$.}
	\label{tab}
	\begin{center}
		\begin{tabular}{|c|c|c|c|c|c|}
\hline
			Country & $a_{0}$ & $a_{1}$ & $a_{2}$& $a_{3}$ & $R^2$\\
\hline
			Canada & 14.725 & 0.547 & -0.013& -0.001& 0.125  \\
\hline
			Australia & -74.207 & 7.656 & 0.003& -0.025& 0.074  \\
\hline
			Mongolia & 79.273 & -5.273 & -1.02&  -0.025& 0.070  \\
\hline
		\end{tabular}
	\end{center}
\end{table}

Thus, we have examined three areas in three different countries chosen from three different continents to reveal that, in all three cases, the population distribution over an area with relatively uniform environmental conditions exhibits a clear spatial periodicity. Having considered the correlation between the population density and the main environmental properties, we have shown that the correlation is very weak and hence the nearly-periodic pattern is unlikely caused by the effect of the environmental factors. Note that the three considered countries are vastly different in term of their average climate, history, ethnicity and culture. This leads us to assume that there can be a generic mechanism resulting in the emergence of the observed spatial pattern. We further assume that this is a dynamical mechanism originated in the nonlinear interaction between the human demography and the distribution of resources or wealth. The corresponding mathematical model is considered in the next section.


\section{Mathematical model}

In order to describe the dynamics of the human population, we use the simple, ``conceptual'' economic-demographic model earlier developed in \cite{Volpert}. The model quantifies the state of the human society at a given location in space $x$ at a given time $t$ by two state variables, the population density $p(x,t)$ and the concentration of wealth $u(x,t)$. Note that, whilst due to its meaning $p\ge 0$, variable $u$ must not necessarily be non-negative; negative values of wealth can be regarded as debt. In the baseline 1D case (which is relevant in case of the population distribution in a narrow stripe, cf.~the examples in the previous section), the model consists of two partial differential equations of reaction-diffusion type:
\begin{eqnarray}
\label{aaz}
\frac{\partial u}{\partial t} &=& D_u\frac{\partial ^2 u}{\partial x^2} + F(p,u),\\[2pt]
\label{bbz}
\frac{\partial p}{\partial t} &=& D_p\frac{\partial ^2 p}{\partial x^2} + G(p,u),
\end{eqnarray}
(where we neglect, for the sake of simplicity, possible effects of cross-diffusion \cite{Volpert}).
Here the first term in the right-hand side of Eq.~(\ref{bbz}) accounts for the population movement in space, that we assume can be considered, at least over certain spatial and temporal scales, as
random \cite{Codling,Turchin98,Viswanathan} (for a detailed discussion of the ``bugbear of randomness'' see \cite{Turchin98})
and can be described mathematically as standard Fickian diffusion. The diffusion term in Eq.~(\ref{aaz}) describes local wealth redistribution due to the economic activities such as trade and investments, and/or taxes.

In order to specify the population growth, we consider the reaction term in Eq.~(\ref{bbz}) in the following form:
\begin{equation}\label{popgrowrate}
G(u,p) = \alpha p\left[K(u)-p\right] - \sigma(u) p,
\end{equation}
where the first term in the right-hand side describes the reproduction rate of the population. We therefore consider it to be the standard logistic population growth with the fertility rate $\alpha$ and the  carrying capacity $K$. The second term describes mortality, $\sigma$ being the mortality rate. Both the carrying capacity and the mortality rates depend on the wealth.
For the mortality rate, we consider it to be a monotonously decreasing function of wealth.
It takes into account the general observation that, on average, the mortality rate is lower for rich people, e.g.~due to access to better health services and/or healthier life style \cite{Deaton13}. In particular, there is evidence that in the USA wealthier people tend to live longer \cite{Chetty16}. More specifically, we consider the following generic Monod-type parametrization:
\begin{equation}\label{sigma}
\sigma(u) = \sigma_0-\frac{\sigma_1 u}{c_0+u},
\end{equation}
where $c_0$, $\sigma_0$ and $\sigma_1$ are positive parameters, $\sigma_0>\sigma_1$.

In order to parameterize $K(u)$, we first recall that it describes an equilibrium population density allowed by the availability of resources \cite{Rees92}. When the resource - in our case, wealth - becomes scarce, the carrying capacity goes to zero, $K(0)=0$. Therefore, for small $u$, $K(u)$ is an increasing function. However, when the resource (wealth) becomes a plenty, $K(u)$ seizes to be monotonic.
There is a certain cultural shift between the low-income and high-income society groups \cite{Small10}.
For people with a low income wealth is the main limiting factor, whilst for people with a high income it is not necessarily so. In particular, low-income people typically live in urban areas, and hence at relatively high population density, e.g.~due to their dependence on public transport and other public services \cite{Glaeser00}. With an increase in income, people tend to move to less densely populated areas such as a rich suburb or a large private estate.
Correspondingly, we consider the carrying capacity $K(u)$ to be an increasing function of wealth for small $u$ but decreasing function for large $u$, tending to a small value (ultimately, to zero) as $u$ tends to infinity.
 More specifically, we consider the carrying capacity in the following form:
\begin{equation}\label{carrcap}
K(u) = \frac{a_2 u}{u^2+c_2^2},
\end{equation}
where $a_2$ and $c_2$ are positive parameters.

In order to specify the reaction term in Eq.~(\ref{aaz}), we first write it as follows:
\begin{equation}\label{F}
F(u,p) = W(u,p)-S(u,p),
\end{equation}
where $W$ and $S$ are the rates of the wealth production and consumption, respectively. Production of wealth is often described by the Cobb-Douglas production function which in the simplest case can be written as \cite{Groth,TheDigEcon}
\begin{equation}\label{CobbDoug}
W = b L^\nu Q^\beta M^\gamma,
\end{equation}
where $L$ is the labour, $Q$ is the capital and $M$ is the available natural resource, a positive coefficient $b$ is a measure of technology, $\nu$, $\beta$ and  $\gamma$ are positive constants \cite{Groth,TheDigEcon}. We assume that the natural resource is not a limiting factor so that $M$ can be kept as constant. We further assume that capital $Q$ is a function of wealth, $Q=f(u)$, and labour is a function of the population density, $L=g(p)$. Equation (\ref{CobbDoug}) then takes the following form:
\begin{equation}\label{production}
W(u, p) = f(u)g(p).
\end{equation}
Due to their meaning, it is reasonable to assume that $f(u)$ and $g(p)$ are increasing functions with saturation. Correspondingly, we choose them in the generic form as the Monod function:
\begin{eqnarray}\label{func-fg}
f(u) = \frac{a_1 u}{u+c_1}, \qquad g(p) = \frac{p}{p+c_2},
\end{eqnarray}
where $a_1$, $c_1$ and $c_2$ are positive parameters.

For the wealth consumption $S$, we assume it to be the result of two processes, i.e.~due to the depreciation (in particular in case of buildings, machinery, etc.) and due to the consumption of the goods and products by the people. For the depreciation, we assume it to be a linear process with a constant rate $a$. The rate of the individual (per capita) consumption, say $c$, can be described by the Keynes linear consumption function, $c = r +sy$, where $y$ is the per capita income and $r$ and $s$ are positive coefficients. Assuming additionally that average income is proportional to the wealth, we arrive at the following expression:
\begin{equation}\label{consumption}
S(u, p) = au + (r + su)p.
\end{equation}

From (\ref{popgrowrate}--\ref{consumption}), we thus obtain the following expressions for the reaction terms:
\begin{eqnarray}
 \label{reactermsF}
F(p,u) &=& \frac{a_1 up}{(u+c_1)(p+c_2)} - \big[au + (r + su)p\big],\\[3pt]
 \label{reactermsG}
G(p,u) &=& \alpha p\left(\frac{a_2 u}{u^2+c_2^2} - p\right) - \left(\sigma_0-\frac{\sigma_1 u}{c_0+u} \right)p.
\end{eqnarray}

\medskip

\section{A glance at the nonspatial system}\label{sec:nonspatial}

We begin with a brief look at the properties of the nonspatial counterpart of the reaction-diffusion system (\ref{aaz}--\ref{bbz}), which is given by the following equations:
\begin{eqnarray}\label{nonspatsys}
\frac{du}{dt} = F(u,p), \qquad \frac{dp}{dt} = G(u,p),
\end{eqnarray}
where functions $G$ and $F$ are given by Eqs.~(\ref{reactermsF}--\ref{reactermsG}). System (\ref{nonspatsys}) was studies in detail in \cite{Zincenko18}. Here we only briefly revisit some of its properties, to the extent that is needed for the goals of this paper.

\begin{figure}[!t]
\centering
 \subfigure[]{\includegraphics[scale=0.5]{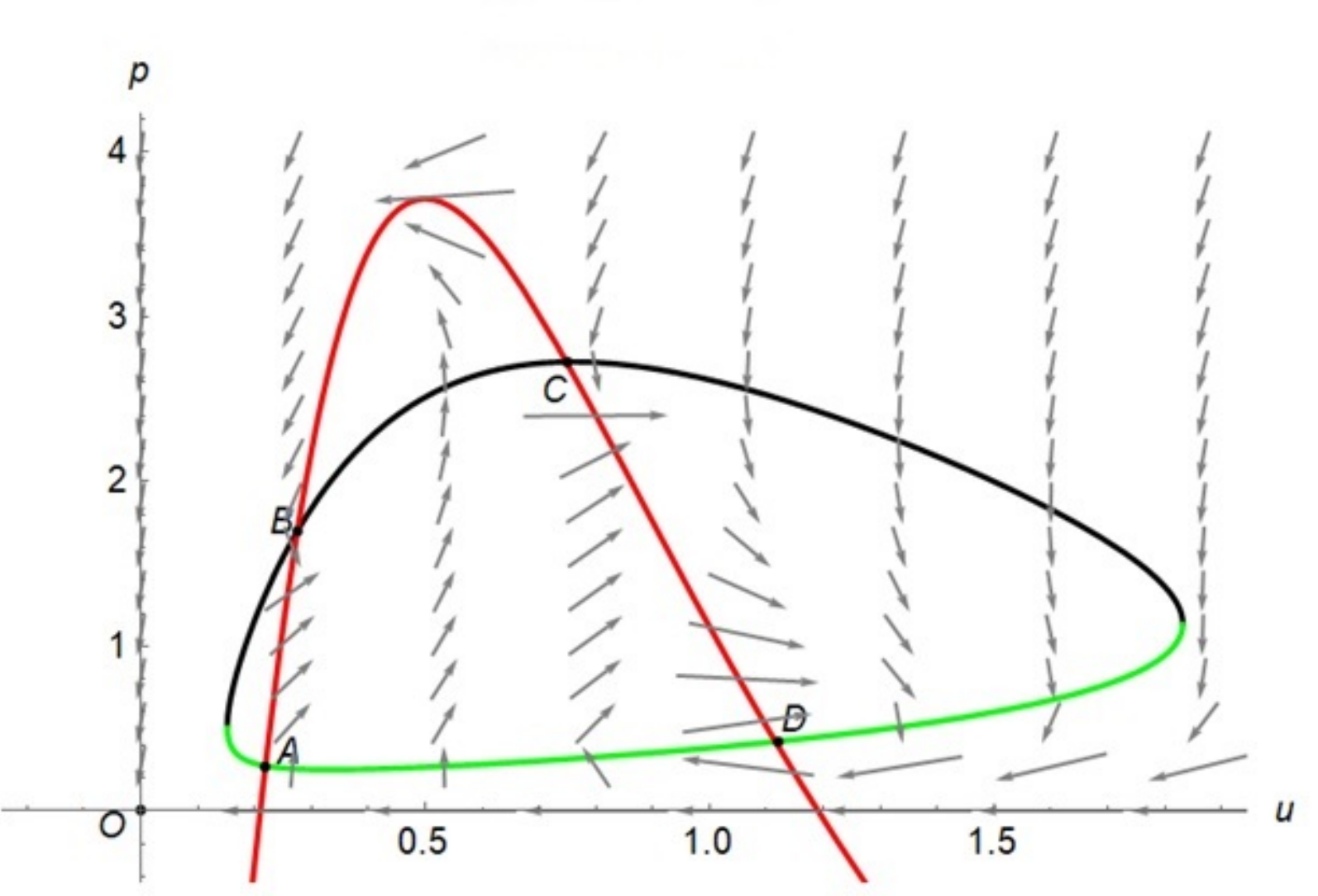}}\hspace{3mm}
 \subfigure[]{\includegraphics[scale=0.42]{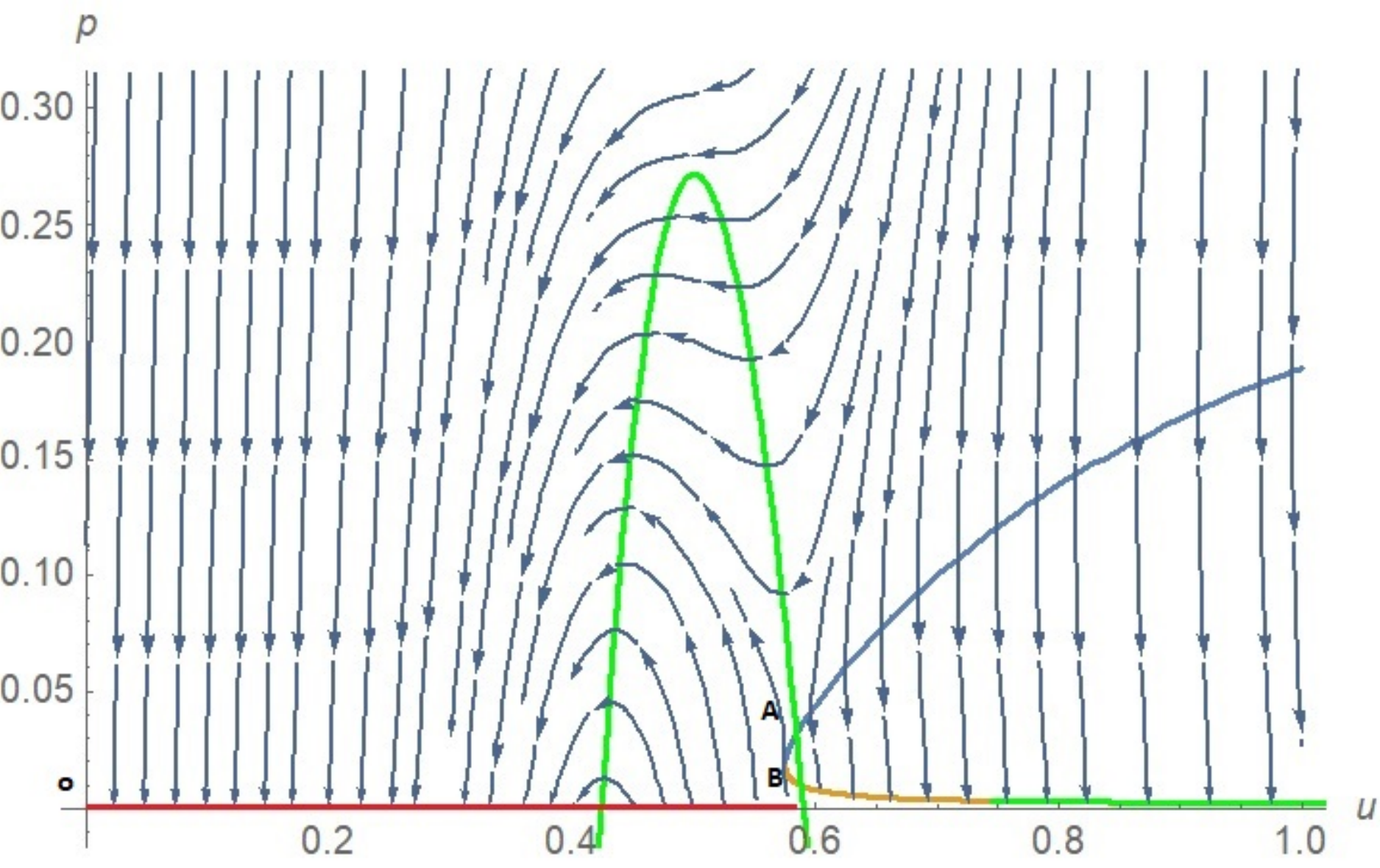}}
\caption{\small Isoclines of the dynamical system (\ref{nonspatsys}) shown for different parameter values, red colour for the $G$-isocline, black-and-green colour for the $F$-isocline.
 (a) The relative position of the isoclines allows for the existence of four positive steady states. Parameters are $c_0=c_1=c_2=1$, $a=15, r=7, s=7.5, a_2=13, c_3=0.5, \sigma_1=0.066, \sigma_0=130, \alpha=14$ and $a_1=150$.
(b) Relative position of the isoclines allowing for the existence of only two positive steady states. Here A is a saddle point and B is a stable focus. Parameters are $c_1=5, c_0=c_2=1,  a=0.01, r=7.5, s=16, a_2=20, c_3=0.5, \sigma_1=0.05, \sigma_0=190, \alpha=9.63$ and $a_1=168$. The origin is always a stable node. Arrows show the direction of the phase flow.}
 \label{r4up1}
\end{figure}
\begin{figure}[!t]
\centering
 \includegraphics[scale=0.43]{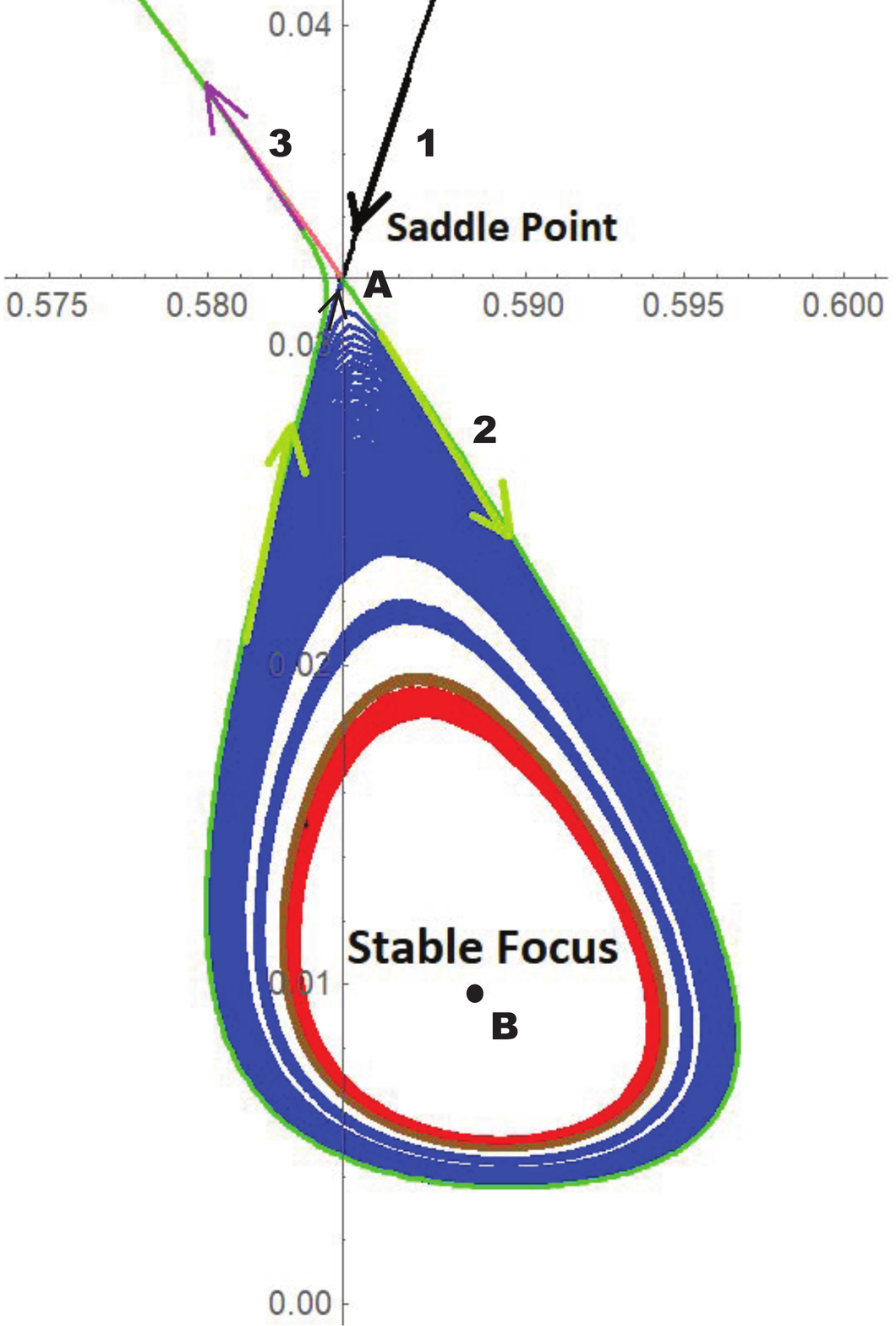}
\caption{\small Fine structure of the phase plane near saddle point $A$ and stable focus $B$. Black curve 1 and the blue curve show the stable manifolds of the saddle. Green curve 2 and magenta line 3 show the unstable manifolds. The brown curve shows the unstable limit cycle. Note that the green curve leaves the vicinity of steady states $A$ and $B$ to eventually go to the stable node $(0,0)$ which is the attractor for the rest of the phase plane (except for the part of the plane inside the limit cycle). The arrows show the direction of the phase flow. The red curve shows a trajectory that starts close to the limit cycle and eventually converges to the stable focus.}
 \label{phg}
\end{figure}

The phase plane of system (\ref{nonspatsys}) is shown in Fig.~\ref{r4up1}.
It is readily seen that the origin is a steady state; a closer look reveals that it is stable node. Inside the first quarter of the phase plane, i.e.~for $u>0$ and $p>0$, the $F$-isocline is a convex closed curve (loop) and the $G$-isocline is an upward-convex, dome-shaped curve. Depending on the relative position of the isoclines (and hence on the parameter values, see \cite{Zincenko18} for details), the number of the positive steady states can be anywhere from 0 to 4. Therefore, in a general case system (\ref{nonspatsys}) can exhibit a rich, multi-stable dynamics and a complicated bifurcation structure where positive states can emerge or disappear.
A typical case corresponding to four positive steady states is shown in Fig.~\ref{r4up1}a.

\begin{figure}[!h]
\begin{tabular}{cc}
\includegraphics[scale=0.31]{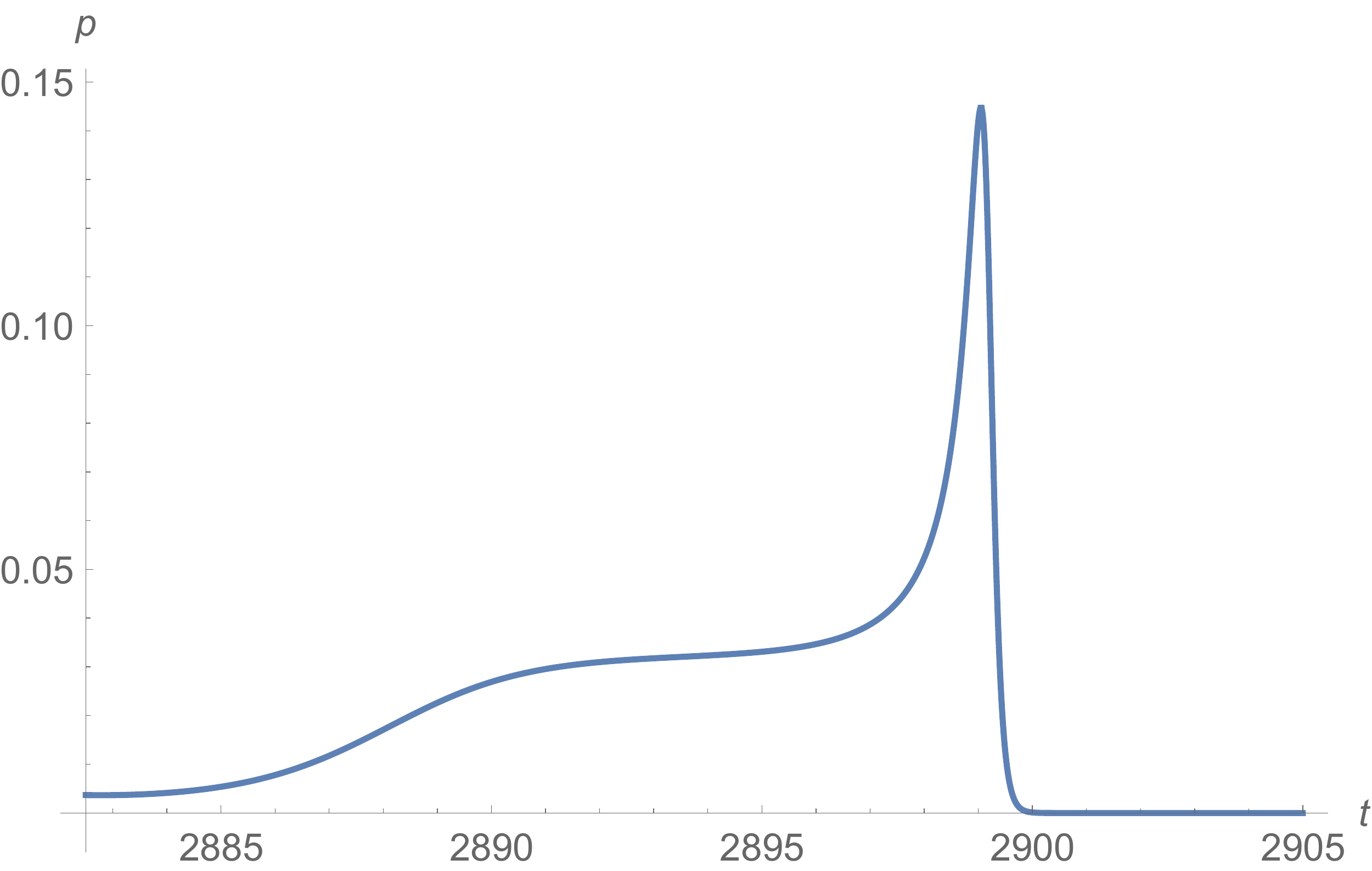} & \includegraphics[scale=0.31]{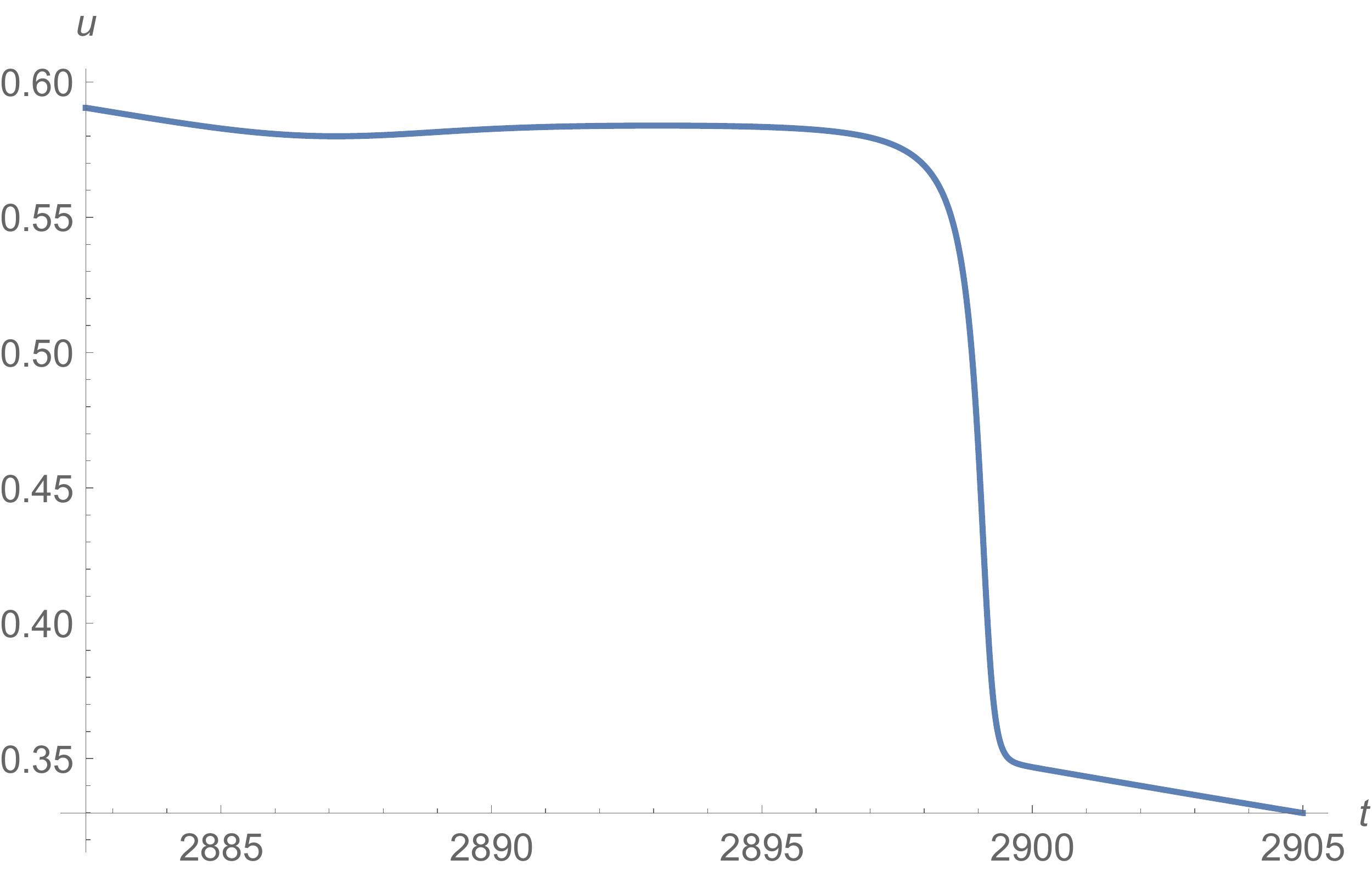}\\
(a) & (b)
\end{tabular}
 \vspace*{1mm}
\begin{tabular}{c}
\includegraphics[scale=0.53]{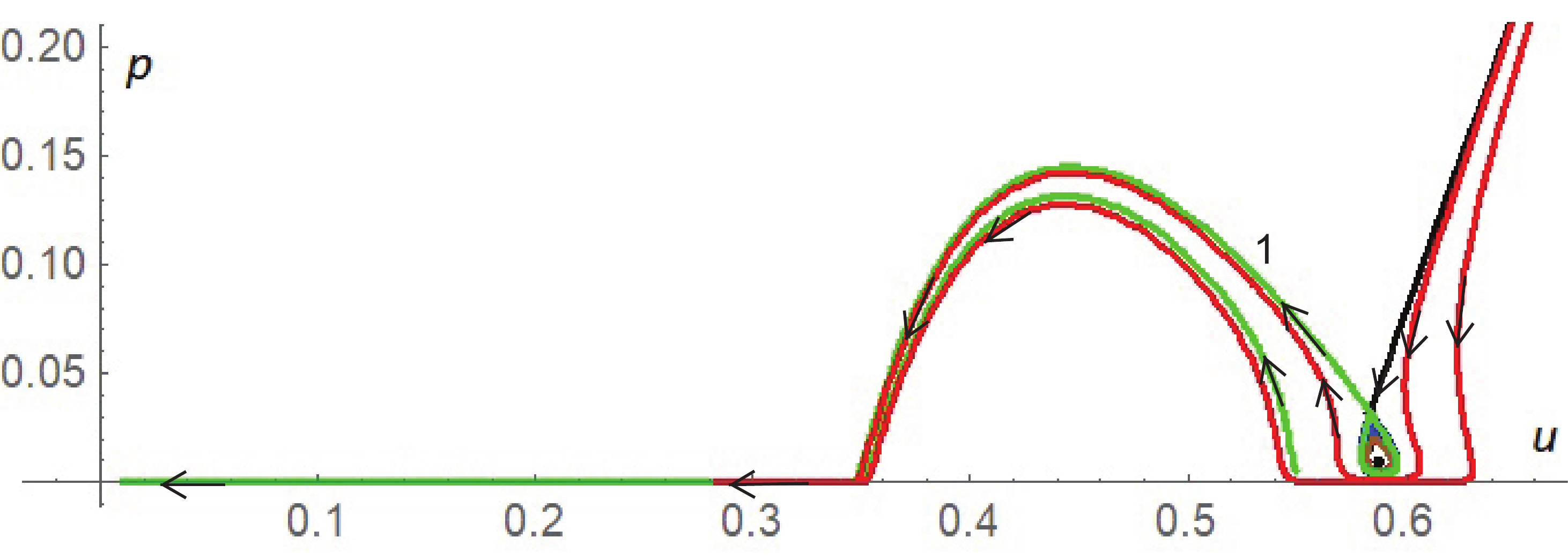}\\
(c)
 \end{tabular}
\caption{\small (a) and (b): population and wealth, respectively, vs time in case the initial conditions correspond to a point in the phase plane outside of the unstable limit cycle (just below the unstable manifold of the saddle point, see curve 2 in Fig.~\ref{phg}). (c) Typical system's trajectories passing through the vicinity of steady states $A$ and $B$, arrows show the direction of the flow. The solution shown in (a,b) corresponds to green curve 1.}
 \label{ut-pt}
\end{figure}

A case where the relative position of the isoclines allows for only two positive steady states is shown in Fig.~\ref{r4up1}b. For these parameters, A is a saddle point and B is a stable focus.
Interestingly, a closer look reveals that even in this case the phase plane has a complicated structure; see Fig.~\ref{phg}.
There are two attractors: the stable node $(0.0)$ and the stable focus $B$, so that the system is bistable.
The attraction basin of stable focus $B$ is bounded by an unstable limit cycle (shown by brown colour). Trajectories that start close to the limit cycle from inside will in the large time limit approach the stable focus; an example is shown by the red curve.
We menton here that the eigenvalues of the system linearized in the vicinity of stable focus $B$ have very small real part (for the parameters of Fig.~\ref{phg}, $\lambda_{1,2}=-0.0000895\pm 0.460312 i$) so that the trajectory approaches the steady state at a very low rate.
Trajectories that start outside of the limit cycle eventually approach the stable node $(0,0)$ except for the special trajectory (the blue curve) that is a part of the stable manifold of saddle point $A$; an example is shown in Fig.~\ref{ut-pt}.

\section{Turing instability conditions}\label{sec:Turing}

We now consider the properties of the spatially-explicit system (\ref{aaz}--\ref{bbz}) with the reaction terms given by (\ref{reactermsF}--\ref{reactermsG}):
\begin{eqnarray}
 \label{spatialu}
\frac{\partial u}{\partial t} &=& D_u\frac{\partial ^2 u}{\partial x^2} + \frac{a_1 up}{(u+c_1)(p+c_2)} - \big[au + (r + su)p\big],\\[2pt]
 \label{spatialp}
\frac{\partial p}{\partial t} &=& D_p\frac{\partial ^2 p}{\partial x^2} + \alpha p\left(\frac{a_2 u}{u^2+c_2^2} - p\right) - \left(\sigma_0-\frac{\sigma_1 u}{1+u} \right)p.
\end{eqnarray}

Equations (\ref{spatialu}--\ref{spatialp}) are complemented with the Neumann `zero-flux' boundary conditions:
\begin{equation}
\label{aaa}
\frac{\partial u}{\partial x}{(0,t)}=0,\quad
\frac{\partial u}{\partial x}{(L,t)}=0,\quad
\frac{\partial p}{\partial x}{(0,t)}=0,\quad
\frac{\partial p}{\partial x}{(L,t)}=0.
\end{equation}

Since our study is motivated by the existence of periodic spatial patterns,
see Section \ref{sec:realdata}, we are particularly interested in the possibility of Turing instability and the corresponding pattern formation. Turing instability is the property of a nonlinear reaction-diffusion system where a steady state that is stable in the corresponding non-spatial system can, under certain parameter constraints, become unstable in the spatial system with respect to periodic perturbation with a certain wavelength \cite{Turing52}.

Let $(\bar{u},\bar{p})$ is a steady state of the nonspatial system and $J$ is the Jacobian matrix evaluated at this steady state:
\begin{equation}
\label{abc00}
\ J=\begin{pmatrix}
\ F_u & \ F_p\\
\ G_u & \ G_p
\end{pmatrix},
\end{equation}
where the subscript denotes the corresponding partial derivative, for instance $F_u = \partial{F(u,p)}/\partial{u}$.
We require that the steady state is stable, so that the following conditions hold:
\begin{eqnarray}
\label{TrDet}
\mbox{(a)}\quad \mbox{tr}(J)<0, \qquad\mbox{and}\qquad\mbox{(b)}\quad \det(J)>0,
\end{eqnarray}
e.g.~see \cite{EdelsteinKeshet05}. In the spatial system, for stability of the corresponding uniform steady state $p(x,t)\equiv\bar{p}$ and $u(x,t)\equiv\bar{u}$ with respect to a periodic perturbation with the wavenumber $k$, conditions (\ref{TrDet}) change to
\begin{eqnarray}
\label{TrDetSpat}
\mbox{(a)}\quad \mbox{tr}(J_k)<0, \qquad\mbox{and}\qquad\mbox{(b)}\quad \det(J_k)>0,
\end{eqnarray}
where
\begin{equation}
\label{abc}
\ J{_k}=J-k^2\begin{pmatrix}
\ D_u&\ 0\\
\ 0&\ D_p
\end{pmatrix}.
\end{equation}
The instability occurs if one of the conditions (\ref{TrDetSpat}) is broken. It is readily seen that condition (\ref{TrDetSpat}a) holds for any $k$. Therefore, the instability can only occur if there is a range of values of $k$ that satisfy the following inequality \cite{EdelsteinKeshet05,Malchow08}:
\begin{equation}
\label{eee}
\min(F(k^2))<0,
\end{equation}
where the characteristic function $F({k^2}) = \det(J_k)$. Taking into account (\ref{abc}), $F(z)$ appears to be a square polynomial, so that inequality (\ref{eee}) is equivalent to \cite{EdelsteinKeshet05,Turing52}:
\begin{equation}
\label{ddd}
D F_u+G_p>2\sqrt{D\det(J)},
\end{equation}
where parameter $D=D_p/D_u$ is the ratio of the diffusion coefficients. In its turn, it appears that a necessary condition for (\ref{ddd}) is that $F_u$ and $G_p$ must be of different sign. Consider $F_u>0$ and $G_p<0$; in this case, $u$ is called the ``activator'' and $p$ the ``inhibitor'' \cite{EdelsteinKeshet05}. Then another necessary condition for (\ref{ddd}) is
$D>D_{cr}>1$ where $D_{cr}$ is a certain critical value that depends on the parameters in the reaction terms \cite{EdelsteinKeshet05,Turing52}.

Now we consider how the generic relation (\ref{ddd}) between the system's feedbacks works in the case of our model (\ref{spatialu}--\ref{spatialp}).
Given the complexity of the bifurcation structure of nonspatial system (\ref{reactermsG}--\ref{nonspatsys}), see Section \ref{sec:nonspatial} and \cite{Zincenko18} for more details, a comprehensive study addressing the Turing instability of all (stable) steady states over the entire parameter range does not seem possible. We therefore concentrate on the specific yet instructive case where there are two positive steady states, a saddle and a stable focus (cf.~Figs.~\ref{r4up1}b and \ref{phg}), in particular to investigate whether the Turing instability may occur for stable steady state $B=(\bar{u},\bar{p})$.

 As a starting point, we consider the following set of parameter values: $a_1=170, c_1=5$, $c_0=c_2=1$, $a=0.01, r=7.5, s=7.5, a_2=20, c_3=0.5, \sigma_1=0.05, \sigma_0=190, \alpha=10$. The corresponding steady state values are $\bar{u}=0.69$ and $\bar{p}=0.00087$ and the Jacobi matrix is
\begin{equation}
J =
\begin{pmatrix}
0.006 & 7.919\\
-0.075 & -0.008
\end{pmatrix}.
\end{equation}
Therefore, at this steady state wealth acts as the activator and population as the inhibitor: the relation similar to the classical resource-consumer system.

\begin{figure}[!b]
\centering
 \includegraphics[scale=0.65]{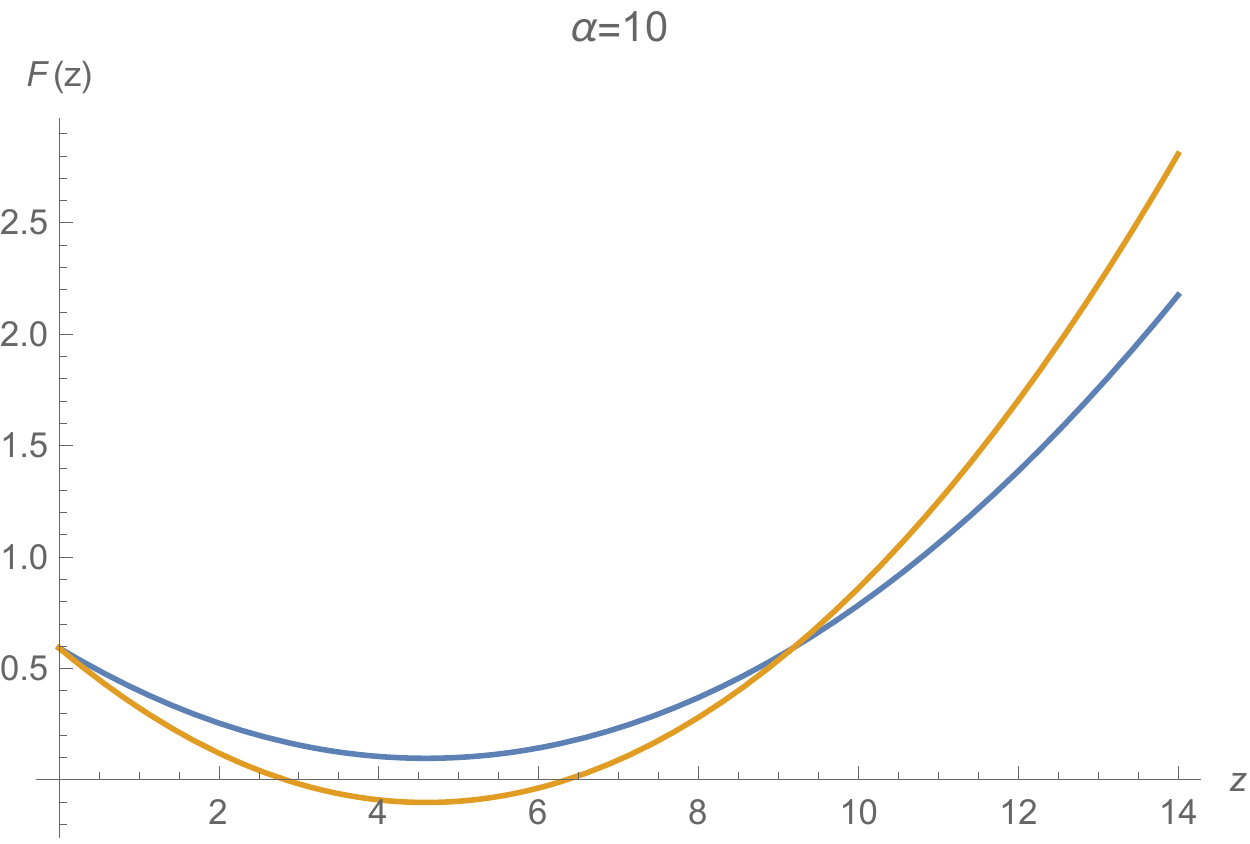}
\caption{\small Examples of the characteristic function $F(z)$ for $D=50000$ (blue curve) and $D=70000$ (orange curve). Other parameters are given in the text; the critical value is $D_{cr}=59746$.}
 \label{Fzz10}	
\end{figure}

For these parameters, the critical value of the diffusivity ratio is readily obtained as $D_{cr}=59746$.
Figure \ref{Fzz10} shows the function $F(z)$ for a subcritical case $\epsilon<\epsilon_{cr}$ where the steady state is stable (as $F(z)>0$ for any $z$ and condition (\ref{TrDetSpat}b) holds for any $k$)
and a supercritical cases $\epsilon>\epsilon_{cr}$ where the steady state is unstable with respect to perturbation with the wavelength from the interval where $F(z)<0$ and hence condition (\ref{TrDetSpat}b) is broken.

\begin{figure}[!t]
\begin{tabular}{cc}
\includegraphics[scale=0.31]{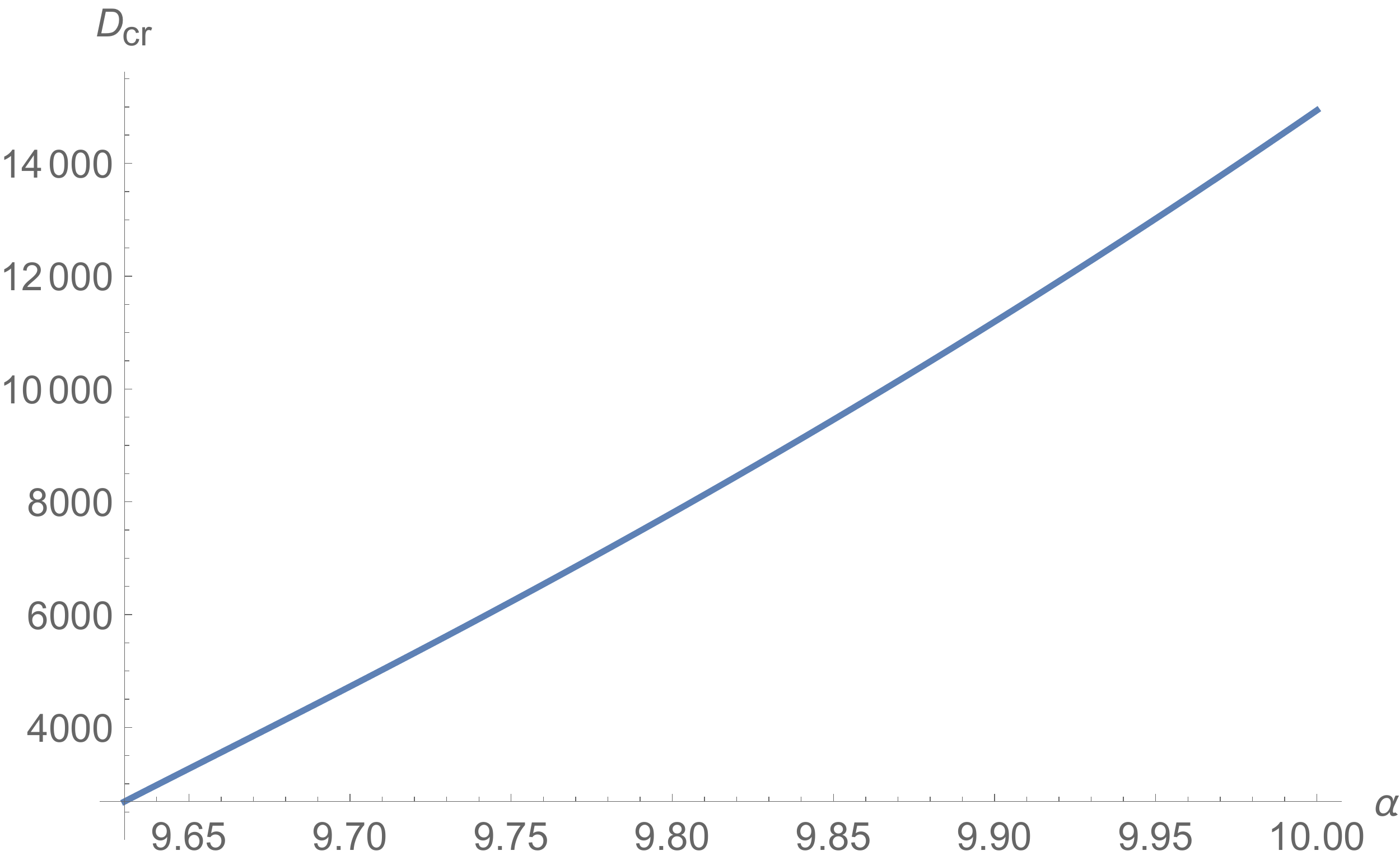} & \includegraphics[scale=0.31]{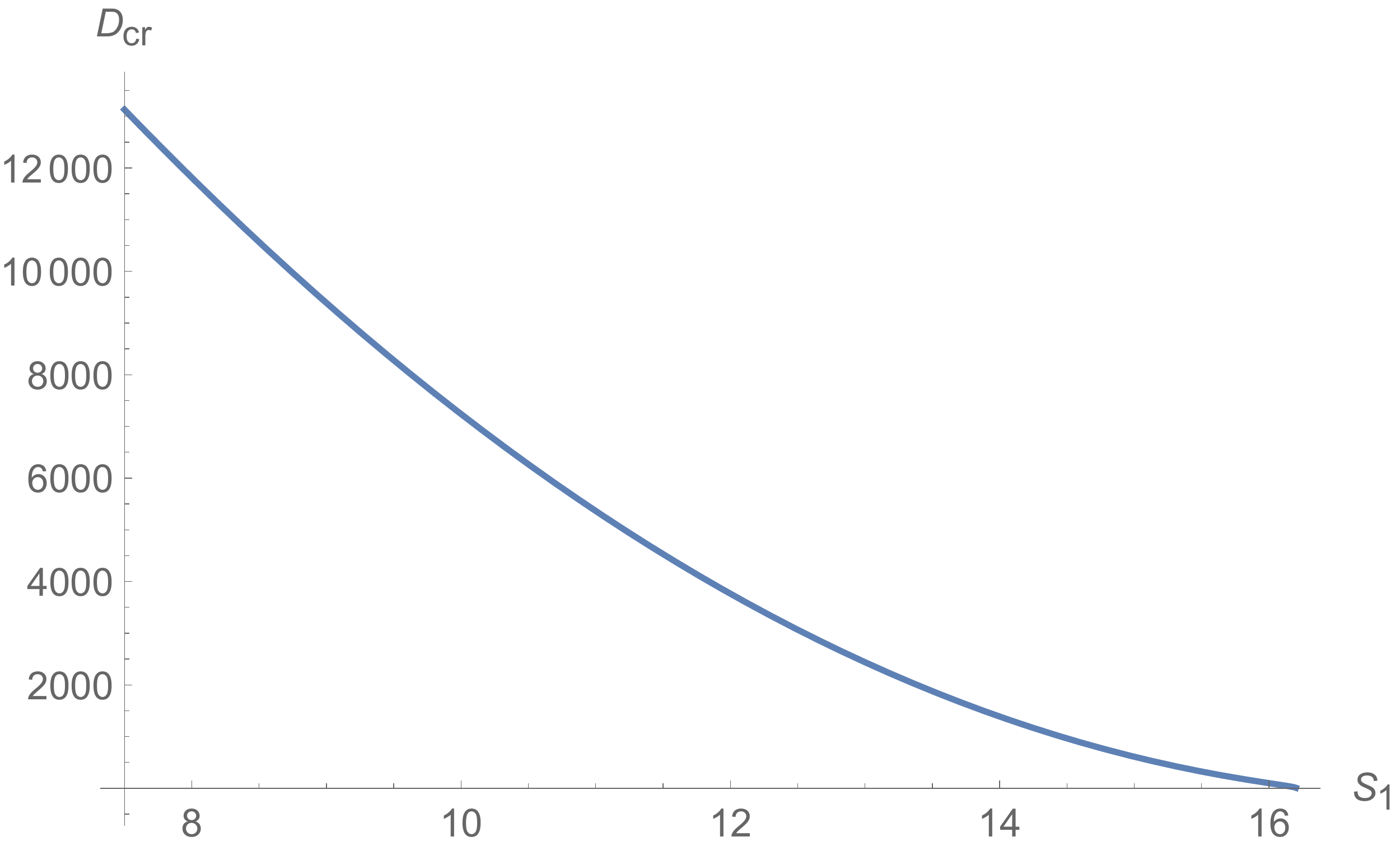}\\
 (a) & (b)
\end{tabular}
\caption{Dependence of the critical diffusivity ratio on model parameters: (a) $D_{cr}$ vs $\alpha$, (b) $D_{cr}$ vs $s$. Other parameters as in Fig.~\ref{Fzz10}.}
 \label{VarAlphaS}
\end{figure}

For the above parameter set, the critical ratio of the diffusion coefficients is very large, which may rise doubts whether it is at all realistic in terms of the real-world dynamics. Therefore, now we are going to consider how the critical relation responds to changes in the parameter values and whether it can be diminished. Indeed, it appears that $D_{cr}$ is rather sensitive with respect to the variation of some of the model parameters; examples are shown in Fig.~\ref{VarAlphaS}. We have found that by varying $\alpha$, $a_1$ and $s$, the critical ratio can be made as small as $D_{cr}=98.5$ (obtained for parameter values $c_1=5$, $c_0=c_2=1$, $a=0.01, r=7.5, s=16, a_2=20, c_3=0.5, \sigma_1=0.05, \sigma_0=190, \alpha=9.63, a_1=168$, the corresponding steady state values are $\bar{u}=0.59$ and $\bar{p}=0.001$).
We mention here that a further reduction of $D_{cr}$ does not appear to be possible: for instance, a further decrease in $\alpha$ (as in Fig.~\ref{VarAlphaS}a) or a further increase in $s$ (as in Fig.~\ref{VarAlphaS}b) make the steady state unstable.

\section{Spatiotemporal dynamics: numerical results}\label{sec:simulations}

In this section, we consider the spatiotemporal dynamics of system (\ref{spatialu}--\ref{spatialp}) that arises as a result of the Turing instability. Note that the fact that the steady state becomes, in a certain parameter range, unstable with regard to spatially heterogeneous perturbations is established analytically (see Section \ref{sec:Turing}) and hence, as such, do not require any confirmation (e.g.~by simulations). However, the mathematical analysis of the instability is based on the system linearization in the vicinity of the steady state and thus is limited to the time interval when the deviation of the solution from the steady state is small. That rises a question what can be the dynamics at the later time, after the deviation from the steady state becomes large enough to be affected by the nonlinearity of the system. Turing instability is known to often lead to the formation of a stationary spatially-periodical pattern \cite{Meinhardt82}, however more complicated dynamics can occur too \cite{Dangelmayr04,Liehr13}.

In order to make an insight into the above question, the reaction-diffusion system (\ref{spatialu}--\ref{spatialp}) is solved numerically by finite-differences using the following initial conditions:
\begin{eqnarray}
u(x,0)=\bar{u}, \qquad p(x,0)=\bar{p}\left(1+0.01\sin\left(\frac{\pi x}{L}\right)\right),
 \label{inicond}
\end{eqnarray}
with the size of the spatial domain $L=120$. The diffusion coefficients are chosen as $D_u=1$ and $D_p=100$, and the values of the reaction parameters are given at the end of the previous section. At the boundaries of the domain, the zero-flux conditions (\ref{aaa}) are used.

\begin{figure}[!t]
 \centering
\subfigure[]{\includegraphics[scale=0.39]{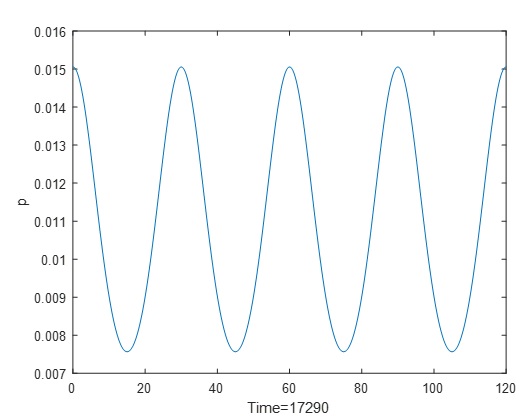}}\hspace{3mm}
\subfigure[]{\includegraphics[scale=0.39]{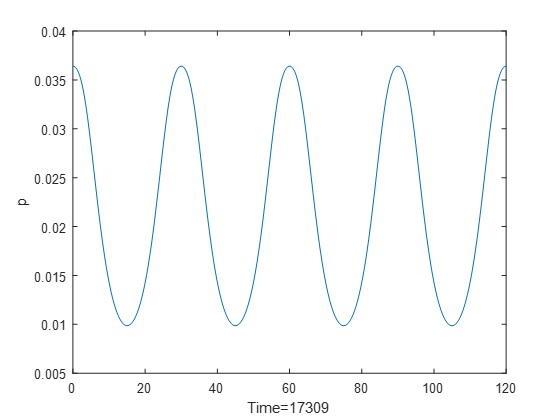}}
\subfigure[]{\includegraphics[scale=0.4]{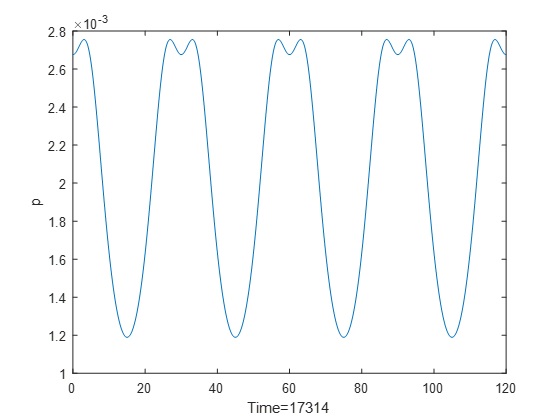}}\subfigure[]{\includegraphics[scale=0.4]{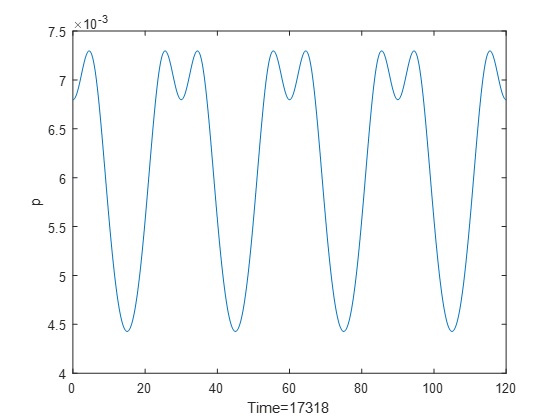}}
\subfigure[]{\includegraphics[scale=0.4]{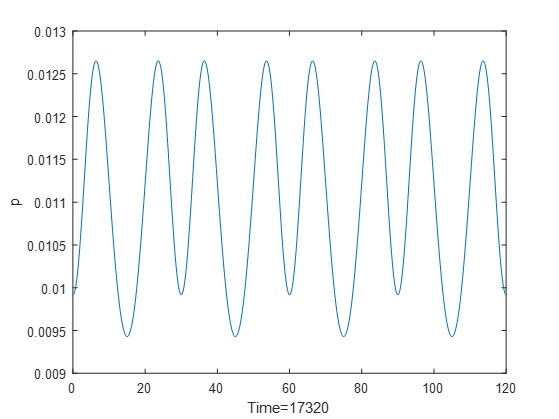}}\subfigure[]{\includegraphics[scale=0.4]{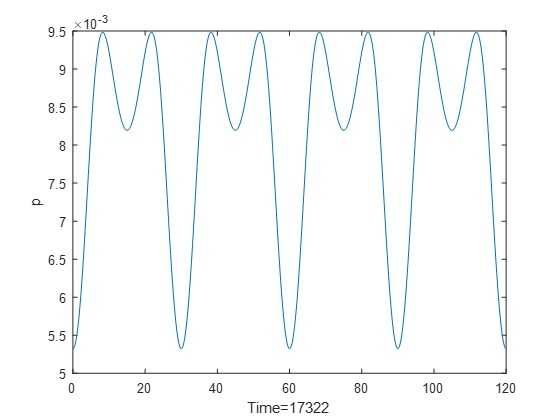}}
\subfigure[]{\includegraphics[scale=0.4]{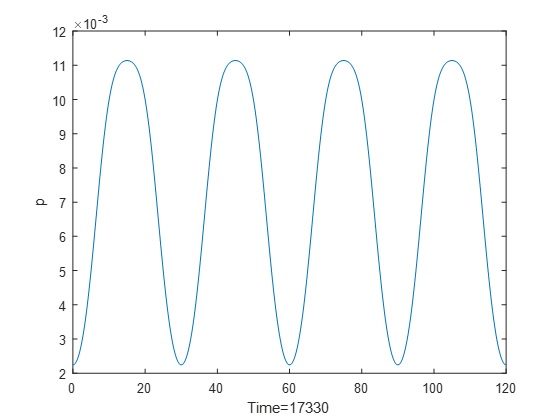}}\subfigure[]{\includegraphics[scale=0.4]{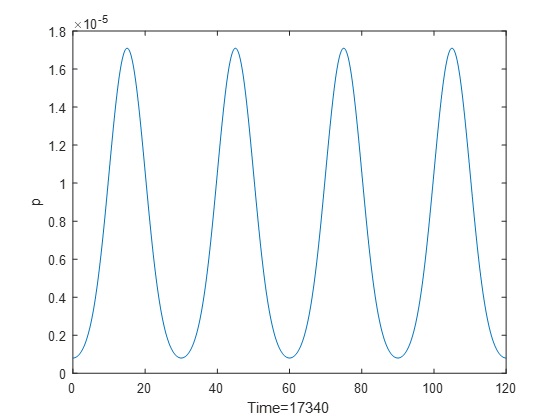}}
\caption{\small Snapshots of the spatial population distribution shown at different moments of time: (a) $t=17290$, (b) $t=17309$, (c) $t=17314$, (d) $t=17318$, (e) $t=17320$, (f) $t=17322$, (g) $t=17330$ and (h) $t=17340$. Note the different order of magnitude for the vertical axis in panel (h).}
 \label{pattern}
\end{figure}

For these parameter values, $D_p/D_u>D_{cr}=98.5$ so that we expect that a small initial perturbation of the steady state leads to the emergence of a spatial pattern. This is indeed seen in the numerical simulations; see Fig.~\ref{pattern}. At an early stage of the dynamics, the initial conditions (\ref{inicond}) fast evolve (over $t\sim 100$, not shown here) to a nearly stationary periodic spatial pattern, which then remains almost unchanged over a considerable time: until $t\approx17200$, see Fig.~\ref{pattern}a). The spatial distribution then starts evolving at a much faster rate, first showing a considerable (nearly three-fold) increase in the spatial variability, see Fig.~\ref{pattern}b, and then developing a higher-frequency spatial mode, cf.~Figs.~\ref{pattern}c-e. The emergence of the higher-frequency mode is accompanied by a gradual decrease in both the population and wealth densities. At a slightly later time, the higher-frequency mode disappears and the spatial distribution again exhibits four distinct peaks but at different locations, so that the peaks and the troughs exchange places, cf.~Figs.~\ref{pattern}a and \ref{pattern}g. Further dynamics lead to the decay in both system components and eventually to system's extinction, see Figs.~\ref{pattern}g-h.

\begin{figure}[!h]
	\centering
\subfigure[]{\includegraphics[scale=0.51]{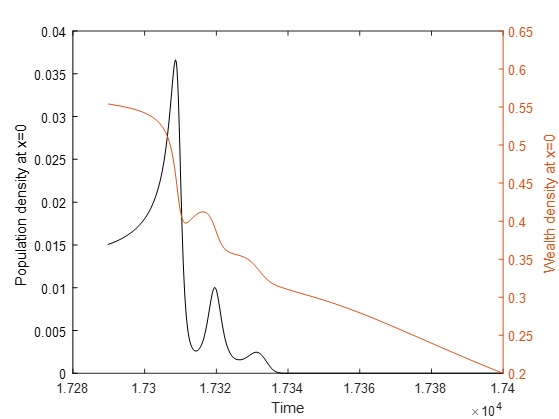}}
\subfigure[]{\includegraphics[scale=0.51]{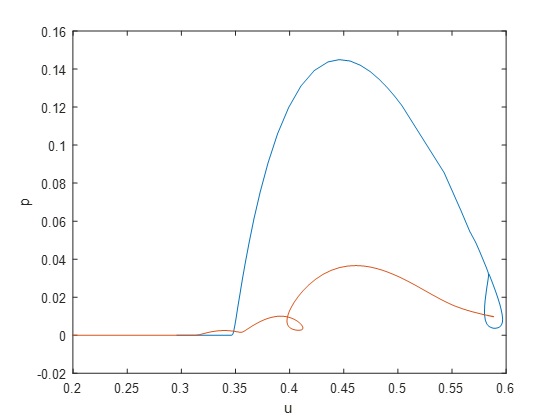}}
\subfigure[]{\includegraphics[scale=0.51]{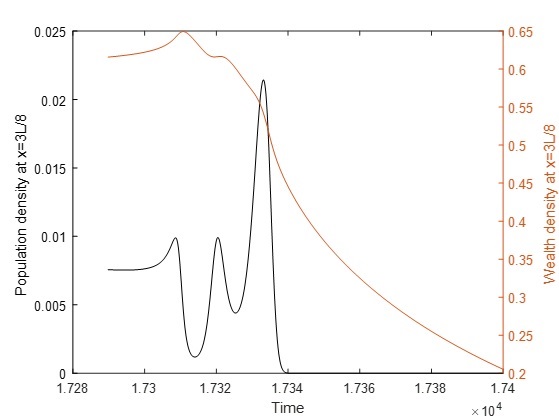}}
\subfigure[]{\includegraphics[scale=0.51]{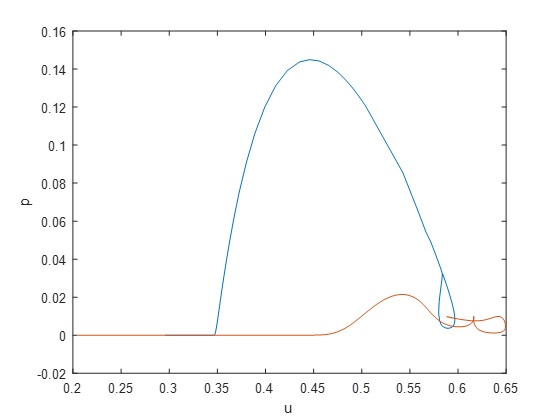}}
\caption{\small (a,c) Figure shows evolution of population distribution and wealth distribution at different spatial points:$x=0$, $x=8L/3$, $x=L$. Change of population density is more steady, apparently due to the fact that the diffusion coefficient for wealth density is far less than the diffusion coefficient for population.
(b,d) Dependence of $p$ and $u$ on time as in (a) and (c) shown as system's
 trajectories in the phase plane, (b) and (d), respectively. The green curve shows the unstable separatrix of the saddle point.}
 \label{TempDep}
\end{figure}

Figure \ref{TempDep} shows, for the same parameter values, the population and wealth densities vs time obtained at two fixed locations in space, i.e.~at the boundary of the domain $x=0$ and at $x=(8/3)L$, see Figs.~\ref{TempDep}a and c, respectively. It is readily seen that the change of the nearly-stationary dynamics (until approximately $t=17280$) to much faster dynamics occurs when the local evolution of the system's variables takes them away from the vicinity of the saddle point (see Fig.~\ref{phg}).

\section{Discussion and concluding remarks}

 In this paper, we have revisited factors and mechanisms affecting the spatial distribution of the human population. Such distributions often exhibit remarkable heterogeneity so that the population density in some areas (e.g.~urban) can be much larger than in others (e.g.~rural).
In the modern perspective, there are many factors that contributes to this difference \cite{Glaeser00}.
However, how the heterogeneity in the population distribution has developed over a longer, historical timescale is not always clear.
 A textbook explanation relates the emergence of densely populated areas to the heterogeneity of the environment. Centuries ago, the humankind was much more exposed to the forces of nature than it is nowadays. Areas with milder climatic conditions would have more likely been selected to establish a settlement. Recall also that the agriculture was the main driver of the economy, its efficiency being to a large extent dependant on the properties of the natural environment.

Convincing as it may sound,
in this paper we endeavoured to challenge the above explanation. We first identified a few areas (selected from different parts of the world) where the environmental properties such as the average annual temperature, annual precipitation and the elevation do not show much variability in space over stretches of thousands of miles. We have revealed that, in spite of this apparent spatial homogeneity of the environment, the population density distribution over those areas is clearly heterogeneous - in fact, in all cases exhibiting a nearly-periodic spatial pattern.

We then ask the following question: can there exist another mechanism leading to the formation of a heterogeneous spatial distribution, even in an approximately homogeneous environment? We hypothesize that an appropriate mechanism can result from the nonlinear interplay between the human population and the resources that support its growth, in the manner of resource-consumer interaction \cite{Murdoch13}.
Indeed, it is well known that, in a system of two or more interacting components, a locally stable steady state can become unstable with respect to a spatially heterogeneous perturbation with a certain wave length: the phenomenon called the Turing instability \cite{Turing52}. As a result, a spatially periodic pattern can arise \cite{Meinhardt82}.

Pattern formation due to the Turing instability is well known for chemical and biological systems \cite{EdelsteinKeshet05,Meinhardt82}.
For demographical systems, however, to the best of our knowledge the Turing instability has never been considered.
In this work we have found this phenomenon within the framework of the demographic-economic model (\ref{spatialu}--\ref{spatialp}) proposed in our earlier work \cite{Volpert,Zincenko18}. Conditions of the Turing instability in terms of the model parameters are found analytically (see Section \ref{sec:Turing}). The spatiotemporal dynamics of the system resulting from the instability, including the development of the periodic spatial pattern, was considered in computer simulations (Section \ref{sec:simulations}).

Interestingly, our simulations reveal that the emerging periodical pattern is not the large-time asymptotics of the system, as it often happens in the case of Turing patterns, but the long-term transient dynamics \cite{Morozov20}. After the pattern emerge (at time $t\sim 10^2$), it remains almost unchanged over long time, on the order of $t=10^4$ (for parameters of Fig.~\ref{pattern}, until $t\approx 17200$). Eventually, this quasi-stationary regime turns into a fast spatiotempotal dynamics where the emergence of a higher-frequency spatial mode is followed by the system collapse. Note that for the parameter values outside of the Turing instability range the spatial system can persist at its positive steady state indefinitely. Therefore, ultimately, the Turing instability drives the system to extinction.


Our study leaves a few open questions.
Firstly, we mention that, although our theoretical findings (such as the formation of the Turing's spatial patterns) are in a qualitative agreement with the real-world data (see Section \ref{sec:realdata}), a direct comparison between theory and data
is hardly possible at this stage of research. Such comparison would required a sufficiently accurate estimate of the value of all model's parameters. This is a challenging and tedious task and will become a focus of a separate study.

Secondly, our model is quite schematic. 
One question that arises here is what can be the effect of different ethnicity and/or race on the population dynamics. Different ethnicity can give rise to a different culture and that can affect the ways how the resources are consumed and the wealth is generated and distributed. Although we do not expect that it will change the system's properties completely - the existence of the nearly-periodic spatial pattern in countries as different as Canada, Australia and Mongolia points out at the universality and robustness of this phenomenon - yet this issue should be addressed more carefully, for instance by using more realistic models.

Another factor that was completely disregarded in our present study is the way how the given country's population has originally emerged.
For instance, Canada’s population mostly emerged as a result of migration from Europe, eventually propagating across Canada from East to West. In Australia, Europeans first settled in and around bays and then diffused over the continent. Finally, Mongolia’s population consists of mainly indigenous population. In mathematical terms, the different ways of how different countries came to be populated may correspond to different initial conditions; that, in its turn, may affect the selection of the emerging spatial mode.

Finally, the effect of the system's spatial dimensionality remains an open question. Recall that our spatially explicit model only include one spatial dimension. Arguably, it was appropriate for considering the population dynamics along a narrow stripe, as was the case identified in our real-world examples. Yet a mode general study on pattern formation in the demographic-economic system should consider it in the more realistic case of two spatial dimensions. That may reveal more complicated patterns and more complicated dynamics. That should be a focus of future research.
\\

\vspace*{5mm}

\end{document}